\PassOptionsToPackage{svgnames}{xcolor}
\documentclass[conference]{IEEEtran}
\IEEEoverridecommandlockouts

\usepackage{cite}
\usepackage{amsmath,amssymb,amsfonts}
\usepackage{graphicx}
\usepackage{textcomp}
\usepackage{xcolor}
\def\BibTeX{{\rm B\kern-.05em{\sc i\kern-.025em b}\kern-.08em
    T\kern-.1667em\lower.7ex\hbox{E}\kern-.125emX}}
\usepackage{csquotes}
\usepackage{amsmath}
\usepackage{bm}
\usepackage[hidelinks]{hyperref}
\usepackage{cleveref}
\usepackage{fontawesome}
\usepackage{graphicx}
\usepackage{glossaries}
\usepackage{booktabs}
\usepackage{multirow}
\usepackage{subcaption}
\usepackage{mathtools} 
\usepackage{url}
\usepackage{listings}
\usepackage{float}
\usepackage{newfloat}
\usepackage{enumitem}
\usepackage{algorithm}
\usepackage{algpseudocode}

\usepackage{pgfplots}    

\crefname{equation}{Equation}{Equations}
\crefname{section}{Section}{Sections}
\crefname{figure}{Figure}{Figures}
\crefname{table}{Table}{Tables}
\crefname{equation}{Equation}{Equations}
\crefname{section}{Section}{Sections}
\crefname{figure}{Figure}{Figures}
\crefname{table}{Table}{Tables}
\crefname{lstlisting}{Listing}{Listings}
\crefname{balgorithm}{Algorithm}{Algorithms}
\crefname{blisting}{Listing}{Listing}

\DeclareFloatingEnvironment[fileext=balgorithm,placement={!ht},name=Algorithm]{balgorithm}
\DeclareFloatingEnvironment[fileext=blisting,placement={!ht},name=Listing]{blisting}
\usetikzlibrary{external}
\DeclareUnicodeCharacter{2212}{−}
\usepgfplotslibrary{groupplots,dateplot}
\usetikzlibrary{shapes,shapes.misc, graphs, patterns,shapes.arrows, arrows.meta, positioning, fit, arrows.meta, graphs, shapes.misc, matrix, quotes, patterns, shapes.arrows, shapes.misc, positioning, calc, fit, bending}

\newacronym[shortplural=LMs, longplural=Language Models]{lm}{LM}{Language Model}
\newacronym[shortplural=KGs, longplural=Knowledge Graphs]{kg}{KG}{Knowledge Graph}
\newacronym[shortplural=GUIs, longplural=Graphical User Interfaces]{gui}{GUI}{Graphical User Interface}
\newacronym{ged}{GED}{Graph Edit Distance}
\newacronym{rag}{RAG}{Retrieval Augmented Generation}
\newacronym{oql}{OQL}{Ontology Query Language}
\newacronym[shortplural=BGPs, longplural=Basic Graph Patterns]{bgp}{BGP}{Basic Graph Pattern}
\newacronym{gbnf}{GBNF}{GGML Backus-Naur Form}
\newacronym{onset}{OnSET}{Ontology and Semantic Exploration Toolkit}
\newacronym{bto}{BTO}{Brainteaser Ontology}
\newacronym{hero}{HERO}{HEREDITARY Ontology}
\newacronym{als}{ALS}{Amyotrophic lateral sclerosis}
\newacronym{ms}{MS}{Multiple sclerosis}
\newacronym{nlp}{NLP}{Natural Language Processing}
\newacronym[shortplural=NLs, longplural=Natural Languages]{nl}{NL}{Natural Language}
\newacronym{sus}{SUS}{System Usability Score}

\newacronym{etl}{ETL}{Extraction, Transform and Load}

\AtBeginDocument{%
  \providecommand\BibTeX{{%
 Bib\TeX}}}

\title{Graph Queries from Natural Language using Constrained Language Models and Visual Editing
\thanks{This work is partially supported by the HEREDITARY Project, as part of the European Union's Horizon Europe research and innovation programme under grant agreement No GA 101137074, the Austrian Science Fund (FWF) 10.55776/COE12, Cluster of Excellence {\href{https://www.bilateral-ai.net/home}{Bilateral Artificial Intelligence}} and the FFG HybridAir project \#FO999902654. 
}
}

\makeatletter
\newcommand{\linebreakand}{%
  \end{@IEEEauthorhalign}
  \hfill\mbox{}\par
  \mbox{}\hfill\begin{@IEEEauthorhalign}
}
\makeatother
\author{
  \IEEEauthorblockN{Benedikt Kantz}
  \IEEEauthorblockA{\textit{Institute of Visual Computing} \\
    \textit{Graz University of Technology}\\
    Graz, Austria \\
    benedikt.kantz@tugraz.at}
  \and
  \IEEEauthorblockN{Kevin Innerebner}
  \IEEEauthorblockA{\textit{Institute of Human-Centred Computing} \\
    \textit{Graz University of Technology}\\
    Graz, Austria \\
    innerebner@tugraz.at}
  \and
  \IEEEauthorblockN{Peter Waldert}
  \IEEEauthorblockA{\textit{Institute of Visual Computing} \\
    \textit{Graz University of Technology}\\
    Graz, Austria \\
    peter.waldert@tugraz.at}
  \linebreakand
  \IEEEauthorblockN{Stefan Lengauer}
  \IEEEauthorblockA{\textit{Institute of Visual Computing} \\
    \textit{Graz University of Technology}\\
    Graz, Austria \\
    s.lengauer@tugraz.at}
  \and
  \IEEEauthorblockN{Elisabeth Lex}
  \IEEEauthorblockA{\textit{Institute of Human-Centred Computing} \\
    \textit{Graz University of Technology}\\
    Graz, Austria \\
    elisabeth.lex@tugraz.at}
  \and
  \IEEEauthorblockN{Tobias Schreck}
  \IEEEauthorblockA{\textit{Institute of Visual Computing} \\
    \textit{Graz University of Technology}\\
    Graz, Austria \\
    tobias.schreck@tugraz.at}
  \centering
}

\begin{document}

  \newcommand{\shortauthors}{Kantz et al.}

  \pgfdeclarelayer{background}
  \pgfdeclarelayer{foreground}
  \pgfsetlayers{background,main,foreground}
  \newcommand{\blockcolor}{CadetBlue}

  \maketitle

  \begin{abstract}
    Querying knowledge bases using ontologies is usually performed using dedicated query languages, question-answering systems, or visual query editors for \glspl{kg}. We propose a novel approach that enables users to query the knowledge graph by specifying prototype graphs using \gls{nl} and visually editing them. This approach enables non-experts to formulate queries without prior knowledge of the ontology and specific query languages. Our approach converts \gls{nl} queries to these prototype graphs by utilizing a two-step constrained \gls{lm} generation based on semantically similar features within an ontology. The resulting prototype graph serves as the building block for further user refinements within a dedicated visual query builder. Our approach consistently generates a valid SPARQL query within the constraints imposed by the ontology, without requiring any additional corrections to the syntax or classes and links used. Unlike related \gls{lm} approaches, which often require multiple iterations to fix invalid syntax, non-existent classes, and non-existent links, our approach achieves this consistently. We evaluate the performance of our system using graph retrieval on synthetic queries, comparing multiple metrics, models, and ontologies. We further validate our system through a preliminary user study. By utilizing our constrained pipeline, we show that the system can perform efficient and accurate retrieval using more efficient models compared to other approaches.
    
  \end{abstract}

  \begin{figure*}[bt]
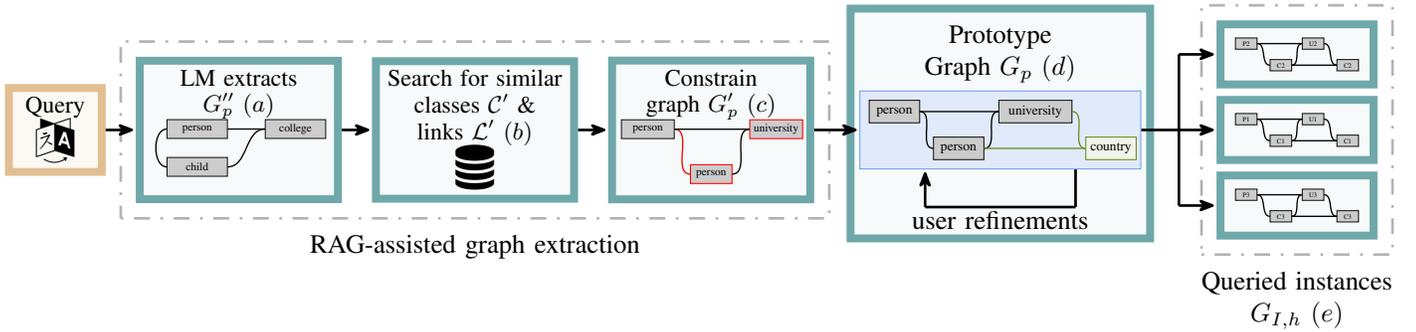

    \centering
\begin{tikzpicture}[
  mapping/.style={rectangle, draw=\blockcolor!90,  fill=\blockcolor!5, minimum size=1cm,  text width=2.4cm, align=center, font={\small},  line width=3pt, minimum height=1.75cm},
  example/.style={mapping, text width=1.8cm, minimum height=0.8cm},
  >={Stealth[round]},
  every new ->/.style={shorten >=1pt, line width=1pt},
  every new --/.style={line width=1pt},
  gray_dotted/.style={
      draw=black!30!white, line width=1pt,
      dash pattern=on 1pt off 4pt on 6pt off 4pt,
      inner sep=2mm, rectangle
    },
  ]
  \node(query)[mapping,  draw=BurlyWood!90,fill=BurlyWood!10, text width=1cm, minimum height=1cm] {Query\\\LARGE\faLanguage};
  \node(erl)[mapping, right=0.4cm of query] {\acrshort{lm} extracts $G_p''$ $(a)$\\ \newcommand{\erlsize}{0.25} \input{figures/erl_unconstrained.tex}};
  \node(candidates)[mapping, right=0.4cm of erl] {Search for similar \\ classes $\mathcal{C}'$ \& links $\mathcal{L}'$ $(b)$\\ \LARGE\faDatabase};
  \node(constrained)[mapping, right=0.4cm of candidates] {Constrain graph $G_p'$ $(c)$\\ \newcommand{\erlsize}{0.25} \input{figures/erl_constrained.tex}};
  \node (desc_query) [gray_dotted,
    fit = (erl) (candidates) (constrained) ] {};
  \node at (desc_query.south) [below, inner sep=2mm] {RAG-assisted graph extraction};
  \node (intermediate) [right=0.6cm of constrained, fill=CornflowerBlue!20, draw=CornflowerBlue!90, text width=3.5cm, align=center] {
    \newcommand{\erlsize}{0.3}
    \input{figures/erl_edit.tex}
  };
  \draw (intermediate.south)+(1,-0.5) coordinate  (intermediate_br);
  \draw (intermediate.south)+(-1,-0.5) coordinate  (intermediate_bl);
  \draw (intermediate.south)+(0,-0.8) coordinate (intermediate_spacer);
  \draw (intermediate.south-|intermediate_br)  coordinate  (intermediate_br_touch);
  \draw (intermediate.south-|intermediate_bl)  coordinate  (intermediate_bl_touch);
  \node (intermediate_lbl) at (intermediate.north) [above, inner sep=1mm, text width=3.25cm, align=center]{
    Prototype Graph $G_p$ $(d)$
  };
  \begin{pgfonlayer}{background}
    \node (intermediate_box) [mapping, fit=(intermediate_lbl) (intermediate) (intermediate_br) (intermediate_bl) (intermediate_spacer)] {};
  \end{pgfonlayer};
  \node (ex_1) [example, right=1cm of intermediate] {
    \newcommand{\erlsize}{0.15}
    \newcommand{\erlperson}{P1}
    \newcommand{\erlchild}{C1}
    \newcommand{\erluni}{U1}
    \newcommand{\erlcountry}{C1}
    \input{figures/erl_ex.tex}
  };
  \node (ex_2) [example, above=0.1cm of ex_1] {
    \newcommand{\erlsize}{0.15}
    \newcommand{\erlperson}{P2}
    \newcommand{\erlchild}{C2}
    \newcommand{\erluni}{U2}
    \newcommand{\erlcountry}{C2}
    \input{figures/erl_ex.tex}
  };
  \node (ex_3) [example, below=0.1cm of ex_1] {
    \newcommand{\erlsize}{0.15}
    \newcommand{\erlperson}{P3}
    \newcommand{\erlchild}{C3}
    \newcommand{\erluni}{U3}
    \newcommand{\erlcountry}{C3}
    \input{figures/erl_ex.tex}
  };
  \draw ($(intermediate.east)!0.5!(ex_1.west)$)  coordinate  (between_ex_1);
  \draw (ex_2.west-|between_ex_1)  coordinate  (between_ex_2);
  \draw (ex_3.west-|between_ex_1)  coordinate  (between_ex_3);
  \node (desc_ex) [gray_dotted,
    fit = (ex_1) (ex_2) (ex_3) ] {};
  \node at (desc_ex.south) [below, inner sep=2mm, text width=4cm, align=center] {Queried instances\\
    $G_{I,h}$ $(e)$};
  \graph [use existing nodes] {
  query->erl->candidates->constrained->intermediate;
  intermediate_br_touch--intermediate_br--["user refinements",inner sep=1pt, auto]intermediate_bl->intermediate_bl_touch;
  intermediate.east--between_ex_1->ex_1.west;
  between_ex_1--{between_ex_2->ex_2.west, between_ex_3->ex_3.west};
  };
\end{tikzpicture}
    \vspace*{-1em}
    \caption{Our query extraction process using \glspl{lm}, using the query example \enquote{a person and the child of a person have the alma mater of the same university}. We transform the \gls{nl} query into a prototype graph using a constrained \gls{lm}. The graph is first approximated using a \gls{lm} $(a)$, where the generated classes and links might not match the ontology yet. This initial guess of the \gls{lm} is used to search for semantically similar relations $(b)$. With the subset of all possible links and nodes, the graph is extracted again and corrected for possible errors $(c)$, resulting in a graph that adheres to the ontology. The resulting graph can be edited $(d)$, e.g., an additional constraint for a country can be added. The resulting prototype graph can be used to perform queries over a \gls{kg} to retrieve instances $(e)$.}
    \label{fig:teaser}
  \end{figure*}

\section{Introduction}
Ontologies are a foundational approach to represent the graph schema for \glspl{kg} and enable knowledge transfers, exploration, and representation for further processing~\cite{Gruber1995Onto}. These ontologies and belonging \glspl{kg} can achieve significant sizes and are, therefore, challenging for users unfamiliar with the knowledge domain to explore and search. One example of such an extensive collection of knowledge is DBpedia~\cite{Lehmann2015DBpediaA}, which compiles linked articles into hundreds of classes and connects them through their various properties within the class hierarchy. Such a rich ontology can pose challenges for users building queries.

The ontologies are usually queried using specialized query languages, such as SPARQL~\cite{Seaborne2024Sparql}, which can pose an additional barrier of entry for users seeking to retrieve structured knowledge from such systems. This paper, therefore, proposes a novel querying strategy, which maps relational queries in \gls{nl} to prototype graph representations, which can then be further adjusted by the user to desired retrieval criteria. The system is tuned for \gls{nl} queries formulated as relations, differing from traditional opaque question-answering approaches, which typically do not visualize the generated queries or their processing, nor allow edits to the query.

Our system extracts the prototype graph structure from the query using the capabilities of \glspl{lm} to extract information from \gls{nl} input -- even in cases where there is no exact match for the required information within the ontology. We achieve a valid prototype graph by constraining the \gls{lm} output using a dynamically created grammar to adhere to the classes and links present within the ontology. This novel addition of the grammar to the query generation by the \gls{lm} enables our system, in turn, to always return prototype graphs that are valid within the context of the used ontology. Our system, furthermore, allows the user to refine and adjust the prototype graph in a visual editor, enabling the correction of mistakes the \gls{lm} might make.

Within this editor, users can refine their queries or extend them to more complex queries, allowing them to edit their queries without modifying the SPARQL query directly or any other intermediate representation in code.

We evaluate the prototype graph extraction performance of our approach using a synthetic benchmark, consisting of sampled sub-graphs and corresponding \gls{lm}-generated \gls{nl} queries. This synthetic benchmark is tested for alignment with human query examples using our synthetic query generation framework. We extend this evaluation to test the integrated system within a user study. This study enables us to compare this combined system of \gls{nl} querying and visual editing with existing visual query editors, as well as formal question-answering benchmarks. Our retrieval approach, furthermore, does not require any metadata about the ontologies or query examples, unlike other systems~\cite{emonet2024llsparql}.

  \section{Related Work}

Previous efforts to map queries presented in \gls{nl} towards complex, constrained query languages~\cite{lei2018ontology, emonet2024llsparql, Ferre2017SPARKLIS} focused on retrieving and generating SPARQL queries directly, resulting in either highly complex systems, or a retrieval process with many iterative refinement steps.

Lei et al.~\cite{lei2018ontology} map a \gls{nl} query onto a particular, tailored, \emph{Ontology Query Language}, used as an intermediate representation, and then converted to SQL.
They use the ontology to find relevant terms in the \gls{nl} query and map them to elements of the ontology.
This approach limits the output to specific instances, hindering both extensibility and exploration.
\textit{SPARKLIS} \cite{Ferre2017SPARKLIS} approaches the mapping from \gls{nl} to a SPARQL query through a mixture of highly constrained \gls{nl} and visual exploration. The constraints placed upon the queries help to adhere to the graph schema but could hinder more straightforward exploration.

More typically, \glspl{kg} are queried using SPARQL~\cite{Seaborne2024Sparql}, which can be generated with \glspl{lm}~\cite{meyer2024assessingsparql,emonet2024llsparql,Liu2024Nov}.
Meyer et al.~\cite{meyer2024assessingsparql} have shown that the out-of-the-box performance of multiple proprietary \glspl{lm} is lacking for direct generation of SPARQL queries from \gls{nl} queries. They demonstrate that adding the ontology, in textual representation, to the query enhances the generation. The generation is further improved if only relevant classes and relations are provided. 
Similarly, Emonet et al.~\cite{emonet2024llsparql} improve upon these results by targeting large-scale federated \glspl{kg}. They develop a \gls{rag} system that automatically augments the \gls{lm} input with relevant ontology classes and manually created example queries.
Finally, Liu et al.~\cite{Liu2024Nov} introduce SPINACH, a question-answering \gls{lm} agent for Wikidata, that iteratively performs actions: searching Wikidata for entities, properties, or example queries, and executing SPARQL queries on demand.
Effectively, these methods highlight that incorporating the ontology is essential to improve generated SPARQL queries.
Nevertheless, these methods require the \glspl{lm} to generate a syntactically and semantically correct SPARQL query, but---due to syntax errors or hallucination of properties---can require feedback error messages to fix the query iteratively.
On the other hand, the visual querying approaches focus on providing approachable systems to facilitate the use of knowledge graphs for non-experts. \emph{GRAPHITE}~\cite{Chau2008GRAPHITE} and \emph{VISAGE}~\cite{Pienta2016VISAGE} both employ visual editing of the prototype graphs to foster the exploration of \glspl{kg} with a focus on fuzzy matching, query suggestions, and fast retrieval times. \emph{RDF Explorer}~\cite{Vargas2019RDF} approaches this task similarly by providing a graph editor, enabling the creation of graph examples to retrieve matching instances from the knowledge graph. The system provides users with query expansion options throughout the application, displaying dynamic results as queries are built. The authors relate their work to previous query builders, notably \emph{Smeagol}~\cite{Clemmer2011}. This alternative follows similar exploration and retrieval paradigms but does not offer a comparable number of SPARQL features. \allowbreak \emph{KGVQL}~\cite{LIU2022108870} defines a novel visual query language to ease the transformation between the visual querying and result set, at the cost of disregarding explorative approaches, favoring the proposed transformation approach to convert between data examples and queries. \emph{Rhizomer}~\cite{GARCIA2022101235} approaches the exploration of knowledge bases by providing different user interfaces to explore only at the top-level, graph-level, or only at the instance level of a single type. \emph{Sparnatural}~\cite{francart2023sparnatural} simplifies many of these exploratory approaches into a tree-based approach.

In comparison to these works, our method utilizes the ontology to constrain the \gls{lm} output, thereby creating a prototype graph that can be directly mapped to a valid SPARQL query. This intermediate graph eliminates syntax errors and hallucinations while enabling the usage of smaller models ($\leq8\text{B}$ parameters) and removing the need for feedback error messages. The system is also robust against semantic mismatches between the query and schema, allowing users to formulate their queries more freely. We also seamlessly integrate this \gls{nl} querying system with a visual interface, enabling users to adjust and refine their query after the mapping phase has been completed.

We, furthermore, did not find any applicable benchmark dataset within the related works analyzed within this paper. The usual system of using a Q\&A system or direct query generation framework is not applicable to our goal of mapping from a \gls{nl} query to a graph representation, which can be evaluated in a more direct way using well-established scores like the $F_1$ score or \gls{ged}~\cite{Abu-Aisheh2015}. The evaluation of our system, therefore, builds upon a synthetic evaluation pipeline, combined with alignment tests and a small-scale user study to confirm our results.

\section{Methodology}

Our \gls{kg} retrieval approach builds on the notion of graph extraction and graph instance retrieval. To realize this notion, we require the graph extraction from \gls{nl} as outlined in \cref{fig:teaser}. The input to our pipeline is a \gls{nl} query, returning a prototype graph $G_p$ with nodes $N_p$ and edges $E_p$. The graph extraction performance is evaluated using a synthetic dataset, generated with our query generation pipeline. The synthetic dataset is shown to be representative of real results through a comparison of using both synthetic and human-written queries on a subset of queries. We provide further details to foster reproducibility in Appendix \cref{ssec:repro}.

\subsection{Graph Extraction}

Our knowledge graph retrieval system builds upon the notion of a prototype graph $G_p \coloneqq (N_p, E_p)$ that serves as the blueprint for all retrieved instances. This graph representation is similar to \gls{bgp}, with the extension of adhering to the schema imposed by the ontology. The sub-graph $G_p$ of the ontology consists of the nodes $N_p$, each one an instance of the classes $\mathcal{C}$, and edges $E_{p}$, each one a link of link types $\mathcal{L}$. The multiset of nodes can contain a class 
multiple times, and a link can only span the allowed end and start types (within the class hierarchy), i.e.
\begin{align*}
  N_p & \coloneqq \big\{n_1, \ldots, n_m\big\} \subseteq \mathcal{C} \,,                                           \\
  E_p & \subseteq \big\{e_i=(n_t, l_j, n_h) \;\big|\; \mathrm{subtypeof}(n_t, \,\mathrm{fromtype}(l_j)),           \\
      & \qquad \mathrm{subtypeof}(n_h,\,\mathrm{totype}(l_j))\big\} \subseteq (N_p\times \mathcal{L}\times N_p)\,.
\end{align*}

We then retrieve the instance graphs $G_{I,h} \coloneqq (O_{I,h}, P_{I,h}) $ from the knowledge base with objects from all vertices or objects $\mathcal{O}$ and predicates from all predicates $\mathcal{P}$ in the \gls{kg} that match the prototype graph $G_p$, i.e.
\begin{align*}
  O_{I,h} & \coloneqq \left\{o_i \in \mathcal{O} \;\big|\; \mathrm{typeof}(o_i) \in N_p \right\} \,,                           \\
  P_{I,h} & \coloneqq \big\{p_{i}=(o_{t}, l_{j},o_{h}) \;\big|\; o_t, o_h \in \mathcal{O}_{I,h}, \mathrm{typeof}(o_t) \in N_p, \\
          & \qquad \mathrm{typeof}(o_h) \in N_p, p_{i} \in \mathcal{P}, l_j \in \mathcal{L} \big\}\,.
\end{align*}

Therefore, this retrieval approach requires a prototype graph $G_p$ that matches the types within the possible classes $\mathcal{C}$ and links $\mathcal{L}$ to provide meaningful results. Our graph extraction pipeline from the \gls{nl} achieves this constrained retrieval using a multistep approach illustrated in \cref{fig:teaser}. This approach first extracts an unconstrained graph $G_p''$ from the \gls{nl} prompt using the structured output of a \gls{lm}~\cite{Liu_2024}, which serves as the basis for retrieval of semantically similar and existing classes $\mathcal{C}$ and links $\mathcal{L}$. These are then used to perform another round of structured generation to get $G_p'$, which is refined to the final prototype graph $G_p$ that can be used to retrieve instance graphs $G_{I,h}$.

\subsubsection{Graph from \gls{nl}}

This first unconstrained graph generation step is required as the basis for further querying of possible classes $\mathcal{C}$ and links $\mathcal{L}$. The \gls{lm} output is nevertheless restrained to return only a specific JSON schema through \gls{gbnf}~\cite{gbnf:online, Liu_2024}. This measure enforces a consistent output that can be parsed and used for further processing. These constraints allow us to construct the intermediate graph $G_p''$, which has no constraints regarding the ontology, i.e., $\mathcal{L}$ and $\mathcal{C}$ are open.

While we could constrain the graph types to the whole ontology at this step, we have no way to enforce the structural correctness of the graph over the outgoing link types, increasing the probability of an invalid graph.

\subsubsection{Constraints from Graph}

The unconstrained graph is used to retrieve candidates for the next generation step. This retrieval is performed for each node $n_{i}\in N_p$ and edge $e_{j}\in E_p$ using sentence embeddings of the node and link description. These embeddings are used to retrieve the top $k$ most similar results in terms of cosine distance from all classes $\mathcal{C}$ and links $\mathcal{L}$. The \gls{lm} generation is then further constrained to only include these results. This retrieval of semantically similar links results in the subset of classes $\mathcal{C}'\subseteq \mathcal{C}$ and links $\mathcal{L}'\subseteq \mathcal{L}$.

\subsubsection{Constrained Graph from \gls{nl}}

Finally, the prototype graph $G_p$ is generated by providing the \gls{lm} with the same instruction as in the first step, but with additional constraints placed upon the output generation through \gls{gbnf}~\cite{gbnf:online}. These use the additional information of the possible candidate classes $\mathcal{C}'$ and links $\mathcal{L}'$ from the previous step. This limited set of only semantically similar classes enables the \gls{lm} still to express any graph from the \gls{nl} query while adhering to the relevant query constraints. The \gls{lm}-generated graph $G_p'$ may contain invalid or flipped links as we cannot enforce a valid graph structure on the output. This problem is mitigated by the previous step of only using the ontology subsets and by cleaning the graph using two rules. The first one exchanges the direction of the edge, essentially swapping the nodes $(n_t, l_j, n_h) \mapsto (n_h, l_j, n_t)$, if the types are flipped, i.e., the link $l_j$ may not go from $n_t$ to $n_h$, but from $n_h$ to $n_t$. Our second rule discards any invalid links from the graph if they violate the type constraints.

\subsection{Synthetic Evaluation Methodology}

The described graph retrieval system is evaluated using synthetically generated queries from a sampled prototype graph $G_{p,s}$. This graph is used to generate queries in \gls{nl} using either a \gls{lm} that is prompted with the graph as input structure or a template-based query generation. We additionally validate our query generation methodology against queries generated by humans by comparing the resulting evaluation metrics. The \gls{nl} prompt is, in turn, used to extract the prototype graph $G_p$ using the method from above. Finally, the sampled and extracted graphs are compared and evaluated for similarity. This evaluation system does not incorporate user refinements for adjusting the output graph, as the methodology evaluates only the directly returned prototype graph $G_p$. We, furthermore, compare the query extraction of our system to the output of the \gls{lm} without any alignments, i.e. the first unconstrained output of our system, in an ablation study performed for all evaluated settings.

\subsubsection{Graph Sampling}
The first step in our synthetic evaluation pipeline is the sampling of prototype graphs $G_{p,s}$ from the ontology. The sampling is based on probabilities derived from the instance counts of the links. We additionally select classes that are lower in the class hierarchy using a similar probabilistic approach. Finally, further links are added based on a random choice of node and a similar probabilistic selection. This probabilistic sampling is only performed for ontologies with sufficient instances, as indicated in~\cref{ssec:eval_models_onto}. The classes and links are sampled uniformly for the other cases. Additional graph sampling details can be found in \cref{ssec:sampling}.

\subsubsection{Generation of the Query}

Next, the \gls{nl} query is generated synthetically from the previously sampled graph $G_{p,s} $ using a \gls{lm}, more specifically the Hermes 3 Llama 3.2 3B model~\cite{teknium2024hermes3technicalreport}. The \gls{lm} is presented with a textual representation of the types of classes and links within the graph and prompted, using a one-shot approach, to generate the queries. We also employ a template-based query generation approach to prevent potential information leaks and evaluate simpler queries~\cite{Sannigrahi_2024}. Additionally, we create 12 human-written queries from the sampled graphs to validate our evaluation methodologies for queries used in practice.

\subsubsection{Scoring the Graphs}

Finally, the prototype graph $G_{p}$ can be extracted from the synthetic query using our graph extraction methodology and compared to the ground truth sampled graph $G_{p,s}$. This evaluation employed two scoring methodologies: one for the retrieval performance of the nodes $N_p$ and relations $E_p$, and another for the graph layout. First, the similarity between the retrieved nodes $N_p$ and sampled nodes $N_{p,s}$ is compared using the $F_1$-score over the sets. This set-based $F_{1,\textit{node}}$-score uses the true positives $TP=|N_p\cap N_{p,s}|$, false positives $FP=|N_p - N_{p,s}|$ and false negatives $FN=|N_{p,s} - N_{p}|$ rates from set intersections and differences. The $F_{1,\textit{rel}}$ of the relations is computed using the same scheme using $E_p$ and $E_{p,s}$.

Second, the graph similarity is computed using the \gls{ged}~\cite{Abu-Aisheh2015}, which employs a stricter notion of similarity, requiring not only isomorphism between the graphs but also the same node and link types for the graphs. To achieve a comparable score to the $F_1$ scores and between different graph sizes, we weigh the distance by the number of nodes and links and invert it, giving the normalized \gls{ged} score
\begin{equation*}
  \textit{GED}_{s}=1-\frac{\textit{GED}}{\max \{|N_p|,|N_{p,s}|\}+\max \{|E_p|,|E_{p,s}|\}}\,.
\end{equation*}

Both evaluation methodologies provide scores for a single query example. We therefore reduce them to single values by averaging the scores over the different queries and evaluation settings.

\subsubsection{Evaluated Models and Ontologies}
\label{ssec:eval_models_onto}
The evaluation of this work relies heavily on the use of \glspl{lm} for graph retrieval, semantic retrieval, and constrained retrieval. We use the Hermes models~\cite{teknium2024hermes3technicalreport} (3B, 8B, and 70B parameter sizes)

for generative retrieval tasks due to their fine-tuned capabilities for structured output, comparing them to two Qwen2.5 (32B parameters) models (Instruct and Code fine-tuned)~\cite{qwen2025qwen25technicalreport}. The Stella 400M~\cite{Zhang2025jasperStella, reimers-2019-sentence-bert} model is used for all semantic retrieval tasks. Further implementation details can be found in \cref{ssec:repro}.

We evaluate our approach on four ontologies, specifically
\begin{itemize}
  \item DBpedia~\cite{Lehmann2015DBpediaA}, an ontology and \gls{kg} containing mapped information from Wikidata;
  \item Yago 4~\cite{Tanon2020YAGO4}, a similarly curated version of Wikidata on the schema.org classes;
  \item \gls{bto}~\cite{2024-brainteaser-ontology}, a smaller ontology and \gls{kg} focusing on brain-related diseases; and
  \item UniProt~\cite{UniProt2025}, a similarly focused ontology and \gls{kg} containing vast amounts of protein sequences and their functional information.
\end{itemize}
Only DBpedia~\cite{Lehmann2015DBpediaA} and Yago 4~\cite{Tanon2020YAGO4} use the probabilistic sampling based on instance count, as the other two ontologies do not provide such a large \gls{kg} to enable good sampling, using the uniform sampling approach instead.

\subsection{Constraining the \gls{lm}}

The \glspl{lm} are constrained to a schema specified to a grammar whenever we retrieve any graph using language generation. To this end, we use \verb|llama-cpp-agent|\footnote{https://llama-cpp-agent.readthedocs.io/}, which can constrain the \gls{lm} to only output a specific model using a predefined grammar. The first prototype graph $G_p''$ uses a static schema and grammar. In the next generation step, the grammar is adapted to the specific, similar relations found utilizing the output of the first round. The updated grammar is then used to constrain the output only to contain valid types and links, which can be used to generate the correct prototype graph $G_p$. The graph $G_p$ can be converted into a SPARQL query if the user requires it for further use, as shown in \cref{ssec:querygen}.

\subsection{User Interface}

\begin{figure}
  \centering
  \includegraphics[width=0.86\linewidth]{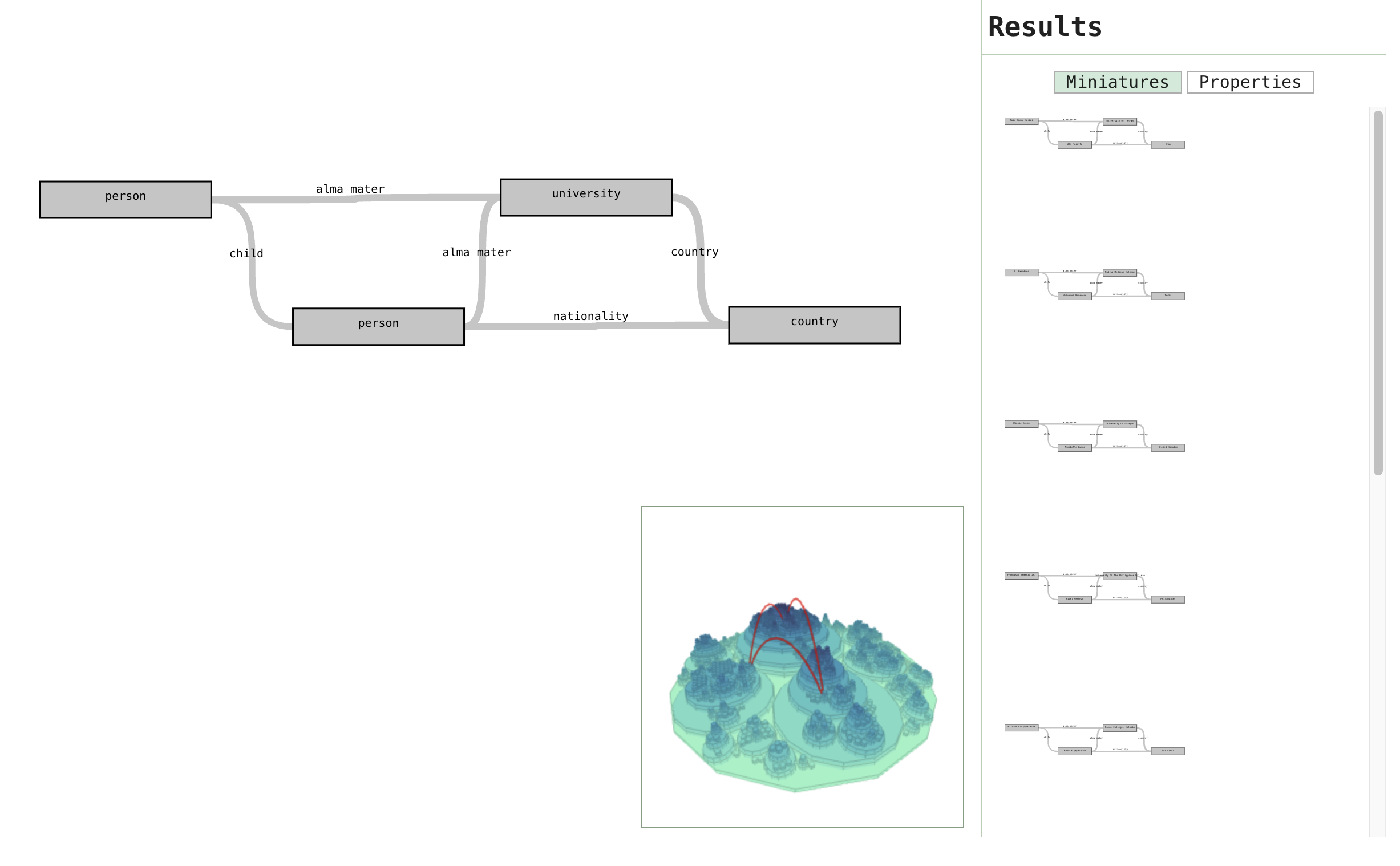}
  \caption{\gls{onset} user interface~\cite{Kantz2025OnSET} showing the query for the example from \cref{fig:teaser}. The user can edit the prototype graph $G_p$ on the left, view the connected nodes at the bottom using a circle-packing visualization of the ontology, and inspect the instances $G_{I,h}$ directly on the right within the interface.}
  \label{fig:user_interface}
\end{figure}

Additionally, we allow the users to refine their initial queries using a node-based editor, where links and nodes can be added to or removed from the graph, all within the constrained link set $\mathcal{L}$, and class set $\mathcal{C}$. This refinement can be helpful if our pipeline fails to find the correct graph or if the users want to refine their search without an additional text prompt. We support a similar refinement process as the second step in our pipeline, which involves performing semantic search for link retrieval. We use the cosine similarity of sentence embeddings~\cite{reimers-2019-sentence-bert} to compare possible new links that the user can search through when adding these. The query editor is based on the prior work within the \gls{onset}~\cite{Kantz2025OnSET} and shown in \cref{fig:user_interface}.

Furthermore, we transparently display how the queries are built using the \gls{lm} through a similar flow to that shown in \cref{fig:teaser}. This visualization should hold our system accountable for any mistakes and errors that occur during processing and may guide the user towards specifying more precise queries or exploring other querying avenues.

\subsection{User Study}

We additionally conduct a preliminary user study to support our claims on usability and improvements in query formulation. Our study focuses on two aspects of the user's experience:
\begin{enumerate}
  \item The user's ability to formulate the correct queries given increasingly complex tasks, with the option to adjust queries using our proposed user interface.
  \item The effectiveness of the integrated \gls{nl} query interface and visual editor.
\end{enumerate}

To validate our claims, we first task the users with completing three increasingly complex queries, as shown in \cref{tab:user_study_tasks}. Each user watches a 2-minute introductory video and has 20 minutes to perform all four tasks. We time the completion of each task and note whether it has the correct number of nodes and links, as well as the correct node types, link types, and constraints. After they complete all tasks or run out of time, the user is presented with a \gls{sus}~\cite{brooke1996sus} evaluation questionnaire to estimate their load on our system. We compare our system to the RDF Explorer, which is evaluated using a similar strategy \cite{Vargas2019RDF}.

\begin{table}
  \centering
  \caption{User tasks on the DBpedia \gls{kg}.}
  \label{tab:user_study_tasks}
  \begin{tabular}{lp{7cm}r}
    \toprule
    No. & Task                                                                                                                                                & $k$ \\
    \midrule
    0   & Find all works where the author is a hockey player.                                                                                                 & 2   \\
    1   & Persons that are authors of a work and are gold medalists in a sports event.                                                                        & 3   \\
    2   & Ships that have their home port in a place in the country United States and persons who have the same place as their death place.                   & 4   \\
    3   & A work that is the opening theme for a TV show is composed by a person and that person has a child. The person has the alma mater of an University. & 5   \\
    \bottomrule
  \end{tabular}
\end{table}

  \section{Results}

We demonstrate our system's capabilities to retrieve the correct graph from the query using our synthetic graph generation and evaluation pipeline. We evaluate the system at three different node sample counts, $k \in \{2,3,5,7\}$, to model varying degrees of complexity in the queries. Each node sample count $k$ is sampled for 128 synthetic queries.
Additionally, we use five different open-weight models to test the dependence on model size and type.
\newcommand{\scoresfigure}[2]{
  \centering
  \begin{subfigure}[c]{0.24\linewidth}
    \input{generated/figures/compare_DBpedia_#1.tex}
    \vspace*{-1.5em}
    \caption{#2 on DBPedia}
  \end{subfigure}
  \begin{subfigure}[c]{0.24\linewidth}
    \input{generated/figures/compare_Yago_#1.tex}
    \vspace*{-1.5em}
    \caption{#2 on Yago}
  \end{subfigure}
  \begin{subfigure}[c]{0.24\linewidth}
    \input{generated/figures/compare_BTO_#1.tex}
    \vspace*{-1.5em}
    \caption{#2 on \acrshort{bto}}
  \end{subfigure}
  \begin{subfigure}[c]{0.24\linewidth}
    \input{generated/figures/compare_UniProt_#1.tex}
    \vspace*{-1.5em}
    \caption{#2 on UniProt}
  \end{subfigure}
}
\begin{figure*}
  
  \centering
  \begin{subfigure}[c]{0.24\linewidth}
\begin{tikzpicture}[
scale=0.5,
every axis/.style={
                    legend pos=south west, 
                    legend style={
                font=\small},
                    legend columns=2, 
                }
]

\definecolor{lightblue161201244}{RGB}{161,201,244}
\definecolor{lightgreen141229161}{RGB}{141,229,161}
\definecolor{lightgrey203}{RGB}{203,203,203}
\definecolor{lightsalmon255159155}{RGB}{255,159,155}
\definecolor{lightsalmon255180130}{RGB}{255,180,130}
\definecolor{thistle208187255}{RGB}{208,187,255}
\definecolor{whitesmoke240}{RGB}{240,240,240}

\begin{axis}[
axis line style={whitesmoke240},
height=1.5\textwidth,
tick align=outside,
tick pos=left,
width=2\textwidth,
x grid style={lightgrey203},
xlabel={\(\displaystyle k\)},
xmajorgrids,
xmin=1.75, xmax=7.25,
xtick style={color=black},
xtick={1,2,3,4,5,6,7,8},
xticklabels={
  \(\displaystyle {1}\),
  \(\displaystyle {2}\),
  \(\displaystyle {3}\),
  \(\displaystyle {4}\),
  \(\displaystyle {5}\),
  \(\displaystyle {6}\),
  \(\displaystyle {7}\),
  \(\displaystyle {8}\)
},
y grid style={lightgrey203},
ylabel={Mean \(\displaystyle F_{1,node}\)},
ymajorgrids,
ymin=0, ymax=1,
ytick style={color=black},
ytick={0,0.2,0.4,0.6,0.8,1},
yticklabels={
  \(\displaystyle {0.0}\),
  \(\displaystyle {0.2}\),
  \(\displaystyle {0.4}\),
  \(\displaystyle {0.6}\),
  \(\displaystyle {0.8}\),
  \(\displaystyle {1.0}\)
}
]
\addplot [thick, lightblue161201244, mark=x, mark size=5, mark options={solid,fill opacity=0}]
table {%
2 0.667940979208585
3 0.74788838612368
5 0.749741797965148
7 0.710023418846948
};
\addplot [thick, lightblue161201244, dashed, mark=x, mark size=5, mark options={solid,fill opacity=0}]
table {%
2 0.683568075117371
3 0.752280399339223
5 0.755312087799397
7 0.716396323143728
};
\addplot [thick, lightsalmon255180130, mark=x, mark size=5, mark options={solid,fill opacity=0}]
table {%
2 0.48341307814992
3 0.705425092411394
5 0.731748621748622
7 0.742349081564768
};
\addplot [thick, lightsalmon255180130, dashed, mark=x, mark size=5, mark options={solid,fill opacity=0}]
table {%
2 0.485805422647528
3 0.714220482713633
5 0.714148444148444
7 0.758754077525704
};
\addplot [thick, lightgreen141229161, mark=x, mark size=5, mark options={solid,fill opacity=0}]
table {%
2 0.635726495726496
3 0.748304132973944
5 0.798049794318451
7 0.781842610940972
};
\addplot [thick, lightgreen141229161, dashed, mark=x, mark size=5, mark options={solid,fill opacity=0}]
table {%
2 0.647692307692308
3 0.769418238993711
5 0.804838445136953
7 0.797848008985907
};
\addplot [thick, lightsalmon255159155, mark=x, mark size=5, mark options={solid,fill opacity=0}]
table {%
2 0.489622641509434
3 0.664746543778802
5 0.702896932127701
7 0.749501242742795
};
\addplot [thick, lightsalmon255159155, dashed, mark=x, mark size=5, mark options={solid,fill opacity=0}]
table {%
2 0.444025157232704
3 0.632762782532368
5 0.685825285825286
7 0.720094562647754
};
\addplot [thick, thistle208187255, mark=x, mark size=5, mark options={solid,fill opacity=0}]
table {%
2 0.513888888888889
3 0.637301587301587
5 0.658574562362441
7 0.670032728008188
};
\addplot [thick, thistle208187255, dashed, mark=x, mark size=5, mark options={solid,fill opacity=0}]
table {%
2 0.454938271604938
3 0.583289845618613
5 0.605194805194805
7 0.611072865101432
};
\end{axis}

\end{tikzpicture}
    \vspace*{-1.5em}
    \caption{$F_{1,\textit{node}}$ on DBPedia}
  \end{subfigure}
  \begin{subfigure}[c]{0.24\linewidth}
\begin{tikzpicture}[
scale=0.5,
every axis/.style={
                    legend pos=south west, 
                    legend style={
                font=\small},
                    legend columns=2, 
                }
]

\definecolor{lightblue161201244}{RGB}{161,201,244}
\definecolor{lightgreen141229161}{RGB}{141,229,161}
\definecolor{lightgrey203}{RGB}{203,203,203}
\definecolor{lightsalmon255159155}{RGB}{255,159,155}
\definecolor{lightsalmon255180130}{RGB}{255,180,130}
\definecolor{thistle208187255}{RGB}{208,187,255}
\definecolor{whitesmoke240}{RGB}{240,240,240}

\begin{axis}[
axis line style={whitesmoke240},
height=1.5\textwidth,
tick align=outside,
tick pos=left,
width=2\textwidth,
x grid style={lightgrey203},
xlabel={\(\displaystyle k\)},
xmajorgrids,
xmin=1.75, xmax=7.25,
xtick style={color=black},
xtick={1,2,3,4,5,6,7,8},
xticklabels={
  \(\displaystyle {1}\),
  \(\displaystyle {2}\),
  \(\displaystyle {3}\),
  \(\displaystyle {4}\),
  \(\displaystyle {5}\),
  \(\displaystyle {6}\),
  \(\displaystyle {7}\),
  \(\displaystyle {8}\)
},
y grid style={lightgrey203},
ylabel={Mean \(\displaystyle F_{1,node}\)},
ymajorgrids,
ymin=0, ymax=1,
ytick style={color=black},
ytick={0,0.2,0.4,0.6,0.8,1},
yticklabels={
  \(\displaystyle {0.0}\),
  \(\displaystyle {0.2}\),
  \(\displaystyle {0.4}\),
  \(\displaystyle {0.6}\),
  \(\displaystyle {0.8}\),
  \(\displaystyle {1.0}\)
}
]
\addplot [thick, lightblue161201244, mark=x, mark size=5, mark options={solid,fill opacity=0}]
table {%
2 0.579376498800959
3 0.62218045112782
5 0.647066787807529
7 0.641831560252613
};
\addplot [thick, lightblue161201244, dashed, mark=x, mark size=5, mark options={solid,fill opacity=0}]
table {%
2 0.540287769784173
3 0.493358395989975
5 0.527071899294121
7 0.551150311676627
};
\addplot [thick, lightsalmon255180130, mark=x, mark size=5, mark options={solid,fill opacity=0}]
table {%
2 0.47673860911271
3 0.646947496947497
5 0.683169269754636
7 0.786010480747323
};
\addplot [thick, lightsalmon255180130, dashed, mark=x, mark size=5, mark options={solid,fill opacity=0}]
table {%
2 0.497122302158273
3 0.605860805860806
5 0.633809347833738
7 0.757411448200922
};
\addplot [thick, lightgreen141229161, mark=x, mark size=5, mark options={solid,fill opacity=0}]
table {%
2 0.748441247002398
3 0.728937728937729
5 0.791044698273614
7 0.874381474381474
};
\addplot [thick, lightgreen141229161, dashed, mark=x, mark size=5, mark options={solid,fill opacity=0}]
table {%
2 0.609832134292566
3 0.716636141636142
5 0.731100158208592
7 0.845823545823546
};
\addplot [thick, lightsalmon255159155, mark=x, mark size=5, mark options={solid,fill opacity=0}]
table {%
2 0.616312056737589
3 0.571904761904762
5 0.679339562876148
7 0.69040959040959
};
\addplot [thick, lightsalmon255159155, dashed, mark=x, mark size=5, mark options={solid,fill opacity=0}]
table {%
2 0.504491725768322
3 0.527916666666667
5 0.643576951503781
7 0.651272151272151
};
\addplot [thick, thistle208187255, mark=x, mark size=5, mark options={solid,fill opacity=0}]
table {%
2 0.621276595744681
3 0.647678571428571
5 0.770918367346939
7 0.705992253360674
};
\addplot [thick, thistle208187255, dashed, mark=x, mark size=5, mark options={solid,fill opacity=0}]
table {%
2 0.520567375886525
3 0.692619047619048
5 0.746678519892806
7 0.72421000052579
};
\end{axis}

\end{tikzpicture}
    \vspace*{-1.5em}
    \caption{$F_{1,\textit{node}}$ on Yago}
  \end{subfigure}
  \begin{subfigure}[c]{0.24\linewidth}
\begin{tikzpicture}[
scale=0.5,
every axis/.style={
                    legend pos=south west, 
                    legend style={
                font=\small},
                    legend columns=2, 
                }
]

\definecolor{lightblue161201244}{RGB}{161,201,244}
\definecolor{lightgreen141229161}{RGB}{141,229,161}
\definecolor{lightgrey203}{RGB}{203,203,203}
\definecolor{lightsalmon255159155}{RGB}{255,159,155}
\definecolor{lightsalmon255180130}{RGB}{255,180,130}
\definecolor{thistle208187255}{RGB}{208,187,255}
\definecolor{whitesmoke240}{RGB}{240,240,240}

\begin{axis}[
axis line style={whitesmoke240},
height=1.5\textwidth,
tick align=outside,
tick pos=left,
width=2\textwidth,
x grid style={lightgrey203},
xlabel={\(\displaystyle k\)},
xmajorgrids,
xmin=1.75, xmax=7.25,
xtick style={color=black},
xtick={1,2,3,4,5,6,7,8},
xticklabels={
  \(\displaystyle {1}\),
  \(\displaystyle {2}\),
  \(\displaystyle {3}\),
  \(\displaystyle {4}\),
  \(\displaystyle {5}\),
  \(\displaystyle {6}\),
  \(\displaystyle {7}\),
  \(\displaystyle {8}\)
},
y grid style={lightgrey203},
ylabel={Mean \(\displaystyle F_{1,node}\)},
ymajorgrids,
ymin=0, ymax=1,
ytick style={color=black},
ytick={0,0.2,0.4,0.6,0.8,1},
yticklabels={
  \(\displaystyle {0.0}\),
  \(\displaystyle {0.2}\),
  \(\displaystyle {0.4}\),
  \(\displaystyle {0.6}\),
  \(\displaystyle {0.8}\),
  \(\displaystyle {1.0}\)
}
]
\addplot [thick, lightblue161201244, mark=x, mark size=5, mark options={solid,fill opacity=0}]
table {%
2 0.277519379844961
3 0.650292397660819
5 0.598030570252793
7 0.724358974358974
};
\addplot [thick, lightblue161201244, dashed, mark=x, mark size=5, mark options={solid,fill opacity=0}]
table {%
2 0.352196382428941
3 0.237802840434419
5 0.231036864370198
7 0.428571428571429
};
\addplot [thick, lightsalmon255180130, mark=x, mark size=5, mark options={solid,fill opacity=0}]
table {%
2 0.181395348837209
3 0.656808688387636
5 0.722652116402116
7 0.643356643356643
};
\addplot [thick, lightsalmon255180130, dashed, mark=x, mark size=5, mark options={solid,fill opacity=0}]
table {%
2 0.00775193798449612
3 0.0467836257309941
5 0.0749158249158249
7 0
};
\addplot [thick, lightgreen141229161, mark=x, mark size=5, mark options={solid,fill opacity=0}]
table {%
2 0.409259259259259
3 0.645454545454546
5 0.628780284043442
7 0.571173271173271
};
\addplot [thick, lightgreen141229161, dashed, mark=x, mark size=5, mark options={solid,fill opacity=0}]
table {%
2 0.20962962962963
3 0.218253968253968
5 0.263157894736842
7 0.0865800865800866
};
\addplot [thick, lightsalmon255159155, mark=x, mark size=5, mark options={solid,fill opacity=0}]
table {%
2 0.326153846153846
3 0.59499596448749
5 0.626617826617827
7 0.615384615384615
};
\addplot [thick, lightsalmon255159155, dashed, mark=x, mark size=5, mark options={solid,fill opacity=0}]
table {%
2 0.0115384615384615
3 0
5 0
7 0
};
\addplot [thick, thistle208187255, mark=x, mark size=5, mark options={solid,fill opacity=0}]
table {%
2 0.185353535353535
3 0.366424535916061
5 0.408018609742748
7 0.385714285714286
};
\addplot [thick, thistle208187255, dashed, mark=x, mark size=5, mark options={solid,fill opacity=0}]
table {%
2 0.0492424242424242
3 0.0677966101694915
5 0.00689655172413793
7 0
};
\end{axis}

\end{tikzpicture}
    \vspace*{-1.5em}
    \caption{$F_{1,\textit{node}}$ on \acrshort{bto}}
  \end{subfigure}
  \begin{subfigure}[c]{0.24\linewidth}
\begin{tikzpicture}[
scale=0.5,
every axis/.style={
                    legend pos=south west, 
                    legend style={
                font=\small},
                    legend columns=2, 
                }
]

\definecolor{lightblue161201244}{RGB}{161,201,244}
\definecolor{lightgreen141229161}{RGB}{141,229,161}
\definecolor{lightgrey203}{RGB}{203,203,203}
\definecolor{lightsalmon255159155}{RGB}{255,159,155}
\definecolor{lightsalmon255180130}{RGB}{255,180,130}
\definecolor{thistle208187255}{RGB}{208,187,255}
\definecolor{whitesmoke240}{RGB}{240,240,240}

\begin{axis}[
axis line style={whitesmoke240},
height=1.5\textwidth,
tick align=outside,
tick pos=left,
width=2\textwidth,
x grid style={lightgrey203},
xlabel={\(\displaystyle k\)},
xmajorgrids,
xmin=1.75, xmax=7.25,
xtick style={color=black},
xtick={1,2,3,4,5,6,7,8},
xticklabels={
  \(\displaystyle {1}\),
  \(\displaystyle {2}\),
  \(\displaystyle {3}\),
  \(\displaystyle {4}\),
  \(\displaystyle {5}\),
  \(\displaystyle {6}\),
  \(\displaystyle {7}\),
  \(\displaystyle {8}\)
},
y grid style={lightgrey203},
ylabel={Mean \(\displaystyle F_{1,node}\)},
ymajorgrids,
ymin=0, ymax=1,
ytick style={color=black},
ytick={0,0.2,0.4,0.6,0.8,1},
yticklabels={
  \(\displaystyle {0.0}\),
  \(\displaystyle {0.2}\),
  \(\displaystyle {0.4}\),
  \(\displaystyle {0.6}\),
  \(\displaystyle {0.8}\),
  \(\displaystyle {1.0}\)
}
]
\addplot [thick, lightblue161201244, mark=x, mark size=5, mark options={solid,fill opacity=0}]
table {%
2 0.60084388185654
3 0.684812030075188
5 0.636242478695309
7 0.668902038132807
};
\addplot [thick, lightblue161201244, dashed, mark=x, mark size=5, mark options={solid,fill opacity=0}]
table {%
2 0.236708860759494
3 0.248771929824561
5 0.238454627133872
7 0.409171597633136
};
\addplot [thick, lightsalmon255180130, mark=x, mark size=5, mark options={solid,fill opacity=0}]
table {%
2 0.493920335429769
3 0.592857142857143
5 0.595861678004535
7 0.607134532134532
};
\addplot [thick, lightsalmon255180130, dashed, mark=x, mark size=5, mark options={solid,fill opacity=0}]
table {%
2 0.0867924528301887
3 0.111776753712238
5 0.224586425479283
7 0.258812615955473
};
\addplot [thick, lightgreen141229161, mark=x, mark size=5, mark options={solid,fill opacity=0}]
table {%
2 0.56
3 0.617152829718275
5 0.608730158730159
7 0.71010101010101
};
\addplot [thick, lightgreen141229161, dashed, mark=x, mark size=5, mark options={solid,fill opacity=0}]
table {%
2 0.13875
3 0.198204936424832
5 0.295004895897753
7 0.428571428571429
};
\addplot [thick, lightsalmon255159155, mark=x, mark size=5, mark options={solid,fill opacity=0}]
table {%
2 0.508251900108578
3 0.490856057522724
5 0.474334554334554
7 0.469478320421717
};
\addplot [thick, lightsalmon255159155, dashed, mark=x, mark size=5, mark options={solid,fill opacity=0}]
table {%
2 0.0641693811074919
3 0.0519061185727852
5 0.0583150183150183
7 0.0479577026746838
};
\addplot [thick, thistle208187255, mark=x, mark size=5, mark options={solid,fill opacity=0}]
table {%
2 0.455765199161426
3 0.511256544502618
5 0.515569985569986
7 0.659051205205051
};
\addplot [thick, thistle208187255, dashed, mark=x, mark size=5, mark options={solid,fill opacity=0}]
table {%
2 0.138364779874214
3 0.152817252555472
5 0.147099567099567
7 0.331665770127309
};
\end{axis}

\end{tikzpicture}
    \vspace*{-1.5em}
    \caption{$F_{1,\textit{node}}$ on UniProt}
  \end{subfigure}

  \centering
  \begin{subfigure}[c]{0.24\linewidth}
\begin{tikzpicture}[
scale=0.5,
every axis/.style={
                    legend pos=south west, 
                    legend style={
                font=\small},
                    legend columns=2, 
                }
]

\definecolor{lightblue161201244}{RGB}{161,201,244}
\definecolor{lightgreen141229161}{RGB}{141,229,161}
\definecolor{lightgrey203}{RGB}{203,203,203}
\definecolor{lightsalmon255159155}{RGB}{255,159,155}
\definecolor{lightsalmon255180130}{RGB}{255,180,130}
\definecolor{thistle208187255}{RGB}{208,187,255}
\definecolor{whitesmoke240}{RGB}{240,240,240}

\begin{axis}[
axis line style={whitesmoke240},
height=1.5\textwidth,
tick align=outside,
tick pos=left,
width=2\textwidth,
x grid style={lightgrey203},
xlabel={\(\displaystyle k\)},
xmajorgrids,
xmin=1.75, xmax=7.25,
xtick style={color=black},
xtick={1,2,3,4,5,6,7,8},
xticklabels={
  \(\displaystyle {1}\),
  \(\displaystyle {2}\),
  \(\displaystyle {3}\),
  \(\displaystyle {4}\),
  \(\displaystyle {5}\),
  \(\displaystyle {6}\),
  \(\displaystyle {7}\),
  \(\displaystyle {8}\)
},
y grid style={lightgrey203},
ylabel={Mean \(\displaystyle F_{1,rel.}\) },
ymajorgrids,
ymin=0, ymax=1,
ytick style={color=black},
ytick={0,0.2,0.4,0.6,0.8,1},
yticklabels={
  \(\displaystyle {0.0}\),
  \(\displaystyle {0.2}\),
  \(\displaystyle {0.4}\),
  \(\displaystyle {0.6}\),
  \(\displaystyle {0.8}\),
  \(\displaystyle {1.0}\)
}
]
\addplot [thick, lightblue161201244, mark=x, mark size=5, mark options={solid,fill opacity=0}]
table {%
2 0.622848200312989
3 0.683559577677225
5 0.742941745226009
7 0.708199425846485
};
\addplot [thick, lightblue161201244, dashed, mark=x, mark size=5, mark options={solid,fill opacity=0}]
table {%
2 0.413145539906103
3 0.404460245636716
5 0.419073368819562
7 0.463825367891112
};
\addplot [thick, lightsalmon255180130, mark=x, mark size=5, mark options={solid,fill opacity=0}]
table {%
2 0.623604465709729
3 0.649771689497717
5 0.72957671957672
7 0.710251331819959
};
\addplot [thick, lightsalmon255180130, dashed, mark=x, mark size=5, mark options={solid,fill opacity=0}]
table {%
2 0.321371610845295
3 0.384170471841705
5 0.402427572427572
7 0.428230157641922
};
\addplot [thick, lightgreen141229161, mark=x, mark size=5, mark options={solid,fill opacity=0}]
table {%
2 0.665811965811966
3 0.691262353998203
5 0.752332852332852
7 0.744226902423624
};
\addplot [thick, lightgreen141229161, dashed, mark=x, mark size=5, mark options={solid,fill opacity=0}]
table {%
2 0.474700854700855
3 0.599505840071878
5 0.601498032095047
7 0.596341539606525
};
\addplot [thick, lightsalmon255159155, mark=x, mark size=5, mark options={solid,fill opacity=0}]
table {%
2 0.657232704402516
3 0.709765196401141
5 0.767806267806268
7 0.765142422589231
};
\addplot [thick, lightsalmon255159155, dashed, mark=x, mark size=5, mark options={solid,fill opacity=0}]
table {%
2 0.388364779874214
3 0.444086021505376
5 0.467806637806638
7 0.529846158569563
};
\addplot [thick, thistle208187255, mark=x, mark size=5, mark options={solid,fill opacity=0}]
table {%
2 0.642746913580247
3 0.707762557077626
5 0.76473063973064
7 0.758713372517054
};
\addplot [thick, thistle208187255, dashed, mark=x, mark size=5, mark options={solid,fill opacity=0}]
table {%
2 0.140432098765432
3 0.211872146118721
5 0.300244872972146
7 0.340086600516048
};
\end{axis}

\end{tikzpicture}
    \vspace*{-1.5em}
    \caption{$F_{1,\textit{rel}}$ on DBPedia}
  \end{subfigure}
  \begin{subfigure}[c]{0.24\linewidth}
\begin{tikzpicture}[
scale=0.5,
every axis/.style={
                    legend pos=south west, 
                    legend style={
                font=\small},
                    legend columns=2, 
                }
]

\definecolor{lightblue161201244}{RGB}{161,201,244}
\definecolor{lightgreen141229161}{RGB}{141,229,161}
\definecolor{lightgrey203}{RGB}{203,203,203}
\definecolor{lightsalmon255159155}{RGB}{255,159,155}
\definecolor{lightsalmon255180130}{RGB}{255,180,130}
\definecolor{thistle208187255}{RGB}{208,187,255}
\definecolor{whitesmoke240}{RGB}{240,240,240}

\begin{axis}[
axis line style={whitesmoke240},
height=1.5\textwidth,
tick align=outside,
tick pos=left,
width=2\textwidth,
x grid style={lightgrey203},
xlabel={\(\displaystyle k\)},
xmajorgrids,
xmin=1.75, xmax=7.25,
xtick style={color=black},
xtick={1,2,3,4,5,6,7,8},
xticklabels={
  \(\displaystyle {1}\),
  \(\displaystyle {2}\),
  \(\displaystyle {3}\),
  \(\displaystyle {4}\),
  \(\displaystyle {5}\),
  \(\displaystyle {6}\),
  \(\displaystyle {7}\),
  \(\displaystyle {8}\)
},
y grid style={lightgrey203},
ylabel={Mean \(\displaystyle F_{1,rel.}\) },
ymajorgrids,
ymin=0, ymax=1,
ytick style={color=black},
ytick={0,0.2,0.4,0.6,0.8,1},
yticklabels={
  \(\displaystyle {0.0}\),
  \(\displaystyle {0.2}\),
  \(\displaystyle {0.4}\),
  \(\displaystyle {0.6}\),
  \(\displaystyle {0.8}\),
  \(\displaystyle {1.0}\)
}
]
\addplot [thick, lightblue161201244, mark=x, mark size=5, mark options={solid,fill opacity=0}]
table {%
2 0.817745803357314
3 0.746491228070175
5 0.730952380952381
7 0.611731981468824
};
\addplot [thick, lightblue161201244, dashed, mark=x, mark size=5, mark options={solid,fill opacity=0}]
table {%
2 0.0407673860911271
3 0.217105263157895
5 0.284832451499118
7 0.365225125751442
};
\addplot [thick, lightsalmon255180130, mark=x, mark size=5, mark options={solid,fill opacity=0}]
table {%
2 0.702637889688249
3 0.738034188034188
5 0.698393341076268
7 0.522727272727273
};
\addplot [thick, lightsalmon255180130, dashed, mark=x, mark size=5, mark options={solid,fill opacity=0}]
table {%
2 0
3 0.0927350427350427
5 0.151567944250871
7 0.104731525784157
};
\addplot [thick, lightgreen141229161, mark=x, mark size=5, mark options={solid,fill opacity=0}]
table {%
2 0.899280575539568
3 0.861111111111111
5 0.828255880665519
7 0.914834714834715
};
\addplot [thick, lightgreen141229161, dashed, mark=x, mark size=5, mark options={solid,fill opacity=0}]
table {%
2 0.105515587529976
3 0.386752136752137
5 0.496175176898069
7 0.594896094896095
};
\addplot [thick, lightsalmon255159155, mark=x, mark size=5, mark options={solid,fill opacity=0}]
table {%
2 0.895981087470449
3 0.84
5 0.852961672473868
7 0.923423423423423
};
\addplot [thick, lightsalmon255159155, dashed, mark=x, mark size=5, mark options={solid,fill opacity=0}]
table {%
2 0.141843971631206
3 0.33625
5 0.31639566395664
7 0.54044604044604
};
\addplot [thick, thistle208187255, mark=x, mark size=5, mark options={solid,fill opacity=0}]
table {%
2 0.796690307328605
3 0.7075
5 0.787112622826909
7 0.8235778841042
};
\addplot [thick, thistle208187255, dashed, mark=x, mark size=5, mark options={solid,fill opacity=0}]
table {%
2 0.0898345153664303
3 0.281666666666667
5 0.289304610733182
7 0.51705312231628
};
\end{axis}

\end{tikzpicture}
    \vspace*{-1.5em}
    \caption{$F_{1,\textit{rel}}$ on Yago}
  \end{subfigure}
  \begin{subfigure}[c]{0.24\linewidth}
\begin{tikzpicture}[
scale=0.5,
every axis/.style={
                    legend pos=south west, 
                    legend style={
                font=\small},
                    legend columns=2, 
                }
]

\definecolor{lightblue161201244}{RGB}{161,201,244}
\definecolor{lightgreen141229161}{RGB}{141,229,161}
\definecolor{lightgrey203}{RGB}{203,203,203}
\definecolor{lightsalmon255159155}{RGB}{255,159,155}
\definecolor{lightsalmon255180130}{RGB}{255,180,130}
\definecolor{thistle208187255}{RGB}{208,187,255}
\definecolor{whitesmoke240}{RGB}{240,240,240}

\begin{axis}[
axis line style={whitesmoke240},
height=1.5\textwidth,
tick align=outside,
tick pos=left,
width=2\textwidth,
x grid style={lightgrey203},
xlabel={\(\displaystyle k\)},
xmajorgrids,
xmin=1.75, xmax=7.25,
xtick style={color=black},
xtick={1,2,3,4,5,6,7,8},
xticklabels={
  \(\displaystyle {1}\),
  \(\displaystyle {2}\),
  \(\displaystyle {3}\),
  \(\displaystyle {4}\),
  \(\displaystyle {5}\),
  \(\displaystyle {6}\),
  \(\displaystyle {7}\),
  \(\displaystyle {8}\)
},
y grid style={lightgrey203},
ylabel={Mean \(\displaystyle F_{1,rel.}\) },
ymajorgrids,
ymin=0, ymax=1,
ytick style={color=black},
ytick={0,0.2,0.4,0.6,0.8,1},
yticklabels={
  \(\displaystyle {0.0}\),
  \(\displaystyle {0.2}\),
  \(\displaystyle {0.4}\),
  \(\displaystyle {0.6}\),
  \(\displaystyle {0.8}\),
  \(\displaystyle {1.0}\)
}
]
\addplot [thick, lightblue161201244, mark=x, mark size=5, mark options={solid,fill opacity=0}]
table {%
2 0.795865633074935
3 0.728070175438597
5 0.67610229276896
7 0.663636363636364
};
\addplot [thick, lightblue161201244, dashed, mark=x, mark size=5, mark options={solid,fill opacity=0}]
table {%
2 0.183462532299742
3 0.243859649122807
5 0.317901234567901
7 0.333333333333333
};
\addplot [thick, lightsalmon255180130, mark=x, mark size=5, mark options={solid,fill opacity=0}]
table {%
2 0.782945736434108
3 0.728654970760234
5 0.838095238095238
7 0.722222222222222
};
\addplot [thick, lightsalmon255180130, dashed, mark=x, mark size=5, mark options={solid,fill opacity=0}]
table {%
2 0.160206718346253
3 0.198245614035088
5 0.377314814814815
7 0.461538461538462
};
\addplot [thick, lightgreen141229161, mark=x, mark size=5, mark options={solid,fill opacity=0}]
table {%
2 0.825925925925926
3 0.76969696969697
5 0.800292397660819
7 0.729614325068871
};
\addplot [thick, lightgreen141229161, dashed, mark=x, mark size=5, mark options={solid,fill opacity=0}]
table {%
2 0.32962962962963
3 0.253030303030303
5 0.309941520467836
7 0.0424242424242424
};
\addplot [thick, lightsalmon255159155, mark=x, mark size=5, mark options={solid,fill opacity=0}]
table {%
2 0.815384615384615
3 0.909039548022599
5 0.809615384615385
7 1
};
\addplot [thick, lightsalmon255159155, dashed, mark=x, mark size=5, mark options={solid,fill opacity=0}]
table {%
2 0.18974358974359
3 0.271186440677966
5 0.346153846153846
7 1
};
\addplot [thick, thistle208187255, mark=x, mark size=5, mark options={solid,fill opacity=0}]
table {%
2 0.823232323232323
3 0.833898305084746
5 0.834975369458128
7 0.9
};
\addplot [thick, thistle208187255, dashed, mark=x, mark size=5, mark options={solid,fill opacity=0}]
table {%
2 0.212121212121212
3 0.271186440677966
5 0.344827586206897
7 0.5
};
\end{axis}

\end{tikzpicture}
    \vspace*{-1.5em}
    \caption{$F_{1,\textit{rel}}$ on \acrshort{bto}}
  \end{subfigure}
  \begin{subfigure}[c]{0.24\linewidth}
\begin{tikzpicture}[
scale=0.5,
every axis/.style={
                    legend pos=south west, 
                    legend style={
                font=\small},
                    legend columns=2, 
                }
]

\definecolor{lightblue161201244}{RGB}{161,201,244}
\definecolor{lightgreen141229161}{RGB}{141,229,161}
\definecolor{lightgrey203}{RGB}{203,203,203}
\definecolor{lightsalmon255159155}{RGB}{255,159,155}
\definecolor{lightsalmon255180130}{RGB}{255,180,130}
\definecolor{thistle208187255}{RGB}{208,187,255}
\definecolor{whitesmoke240}{RGB}{240,240,240}

\begin{axis}[
axis line style={whitesmoke240},
height=1.5\textwidth,
tick align=outside,
tick pos=left,
width=2\textwidth,
x grid style={lightgrey203},
xlabel={\(\displaystyle k\)},
xmajorgrids,
xmin=1.75, xmax=7.25,
xtick style={color=black},
xtick={1,2,3,4,5,6,7,8},
xticklabels={
  \(\displaystyle {1}\),
  \(\displaystyle {2}\),
  \(\displaystyle {3}\),
  \(\displaystyle {4}\),
  \(\displaystyle {5}\),
  \(\displaystyle {6}\),
  \(\displaystyle {7}\),
  \(\displaystyle {8}\)
},
y grid style={lightgrey203},
ylabel={Mean \(\displaystyle F_{1,rel.}\) },
ymajorgrids,
ymin=0, ymax=1,
ytick style={color=black},
ytick={0,0.2,0.4,0.6,0.8,1},
yticklabels={
  \(\displaystyle {0.0}\),
  \(\displaystyle {0.2}\),
  \(\displaystyle {0.4}\),
  \(\displaystyle {0.6}\),
  \(\displaystyle {0.8}\),
  \(\displaystyle {1.0}\)
}
]
\addplot [thick, lightblue161201244, mark=x, mark size=5, mark options={solid,fill opacity=0}]
table {%
2 0.854430379746835
3 0.908070175438596
5 0.823045822102426
7 0.814918414918415
};
\addplot [thick, lightblue161201244, dashed, mark=x, mark size=5, mark options={solid,fill opacity=0}]
table {%
2 0.208227848101266
3 0.274210526315789
5 0.295297993411201
7 0.350815850815851
};
\addplot [thick, lightsalmon255180130, mark=x, mark size=5, mark options={solid,fill opacity=0}]
table {%
2 0.852201257861635
3 0.843906810035842
5 0.86109693877551
7 0.840981240981241
};
\addplot [thick, lightsalmon255180130, dashed, mark=x, mark size=5, mark options={solid,fill opacity=0}]
table {%
2 0.223270440251572
3 0.213184843830005
5 0.356221655328798
7 0.458874458874459
};
\addplot [thick, lightgreen141229161, mark=x, mark size=5, mark options={solid,fill opacity=0}]
table {%
2 0.88125
3 0.90977312390925
5 0.857142857142857
7 0.856204906204906
};
\addplot [thick, lightgreen141229161, dashed, mark=x, mark size=5, mark options={solid,fill opacity=0}]
table {%
2 0.257291666666667
3 0.393891797556719
5 0.516510770975057
7 0.539033189033189
};
\addplot [thick, lightsalmon255159155, mark=x, mark size=5, mark options={solid,fill opacity=0}]
table {%
2 0.883821932681867
3 0.881454348121015
5 0.881550671550672
7 0.841870864984072
};
\addplot [thick, lightsalmon255159155, dashed, mark=x, mark size=5, mark options={solid,fill opacity=0}]
table {%
2 0.311617806731813
3 0.383143399810066
5 0.459271469271469
7 0.354731955675352
};
\addplot [thick, thistle208187255, mark=x, mark size=5, mark options={solid,fill opacity=0}]
table {%
2 0.870020964360587
3 0.859162303664921
5 0.867316017316017
7 0.872882672882673
};
\addplot [thick, thistle208187255, dashed, mark=x, mark size=5, mark options={solid,fill opacity=0}]
table {%
2 0.270440251572327
3 0.241012216404887
5 0.277489177489178
7 0.26993006993007
};
\end{axis}

\end{tikzpicture}
    \vspace*{-1.5em}
    \caption{$F_{1,\textit{rel}}$ on UniProt}
  \end{subfigure}

  \centering
  \begin{subfigure}[c]{0.24\linewidth}
\begin{tikzpicture}[
scale=0.5,
every axis/.style={
                    legend pos=south west, 
                    legend style={
                font=\small},
                    legend columns=2, 
                }
]

\definecolor{lightblue161201244}{RGB}{161,201,244}
\definecolor{lightgreen141229161}{RGB}{141,229,161}
\definecolor{lightgrey203}{RGB}{203,203,203}
\definecolor{lightsalmon255159155}{RGB}{255,159,155}
\definecolor{lightsalmon255180130}{RGB}{255,180,130}
\definecolor{thistle208187255}{RGB}{208,187,255}
\definecolor{whitesmoke240}{RGB}{240,240,240}

\begin{axis}[
axis line style={whitesmoke240},
height=1.5\textwidth,
tick align=outside,
tick pos=left,
width=2\textwidth,
x grid style={lightgrey203},
xlabel={\(\displaystyle k\)},
xmajorgrids,
xmin=1.75, xmax=7.25,
xtick style={color=black},
xtick={1,2,3,4,5,6,7,8},
xticklabels={
  \(\displaystyle {1}\),
  \(\displaystyle {2}\),
  \(\displaystyle {3}\),
  \(\displaystyle {4}\),
  \(\displaystyle {5}\),
  \(\displaystyle {6}\),
  \(\displaystyle {7}\),
  \(\displaystyle {8}\)
},
y grid style={lightgrey203},
ylabel={Mean \(\displaystyle GED_{s}\)},
ymajorgrids,
ymin=0, ymax=1,
ytick style={color=black},
ytick={0,0.2,0.4,0.6,0.8,1},
yticklabels={
  \(\displaystyle {0.0}\),
  \(\displaystyle {0.2}\),
  \(\displaystyle {0.4}\),
  \(\displaystyle {0.6}\),
  \(\displaystyle {0.8}\),
  \(\displaystyle {1.0}\)
}
]
\addplot [thick, lightblue161201244, mark=x, mark size=5, mark options={solid,fill opacity=0}]
table {%
2 0.565157801360269
3 0.522562125793359
5 0.449890382989641
7 0.358610575678126
};
\addplot [thick, lightblue161201244, dashed, mark=x, mark size=5, mark options={solid,fill opacity=0}]
table {%
2 0.508186153960802
3 0.433730974907445
5 0.353478284646924
7 0.299329744906976
};
\addplot [thick, lightsalmon255180130, mark=x, mark size=5, mark options={solid,fill opacity=0}]
table {%
2 0.474248913480056
3 0.508626352414812
5 0.457436513746464
7 0.408297940224147
};
\addplot [thick, lightsalmon255180130, dashed, mark=x, mark size=5, mark options={solid,fill opacity=0}]
table {%
2 0.377773980405559
3 0.422680819941094
5 0.353121342793288
7 0.330067369643131
};
\addplot [thick, lightgreen141229161, mark=x, mark size=5, mark options={solid,fill opacity=0}]
table {%
2 0.532675681806117
3 0.52919706670627
5 0.475423529729025
7 0.424784544606753
};
\addplot [thick, lightgreen141229161, dashed, mark=x, mark size=5, mark options={solid,fill opacity=0}]
table {%
2 0.492755392755393
3 0.551899489399489
5 0.479897088383295
7 0.424576709811291
};
\addplot [thick, lightsalmon255159155, mark=x, mark size=5, mark options={solid,fill opacity=0}]
table {%
2 0.50473944294699
3 0.539717284763368
5 0.477271913837681
7 0.436395578901589
};
\addplot [thick, lightsalmon255159155, dashed, mark=x, mark size=5, mark options={solid,fill opacity=0}]
table {%
2 0.391520664869722
3 0.433807628623297
5 0.378433315050962
7 0.397142887535484
};
\addplot [thick, thistle208187255, mark=x, mark size=5, mark options={solid,fill opacity=0}]
table {%
2 0.511342592592593
3 0.527027771512428
5 0.490923638650911
7 0.464549949244578
};
\addplot [thick, thistle208187255, dashed, mark=x, mark size=5, mark options={solid,fill opacity=0}]
table {%
2 0.313161375661376
3 0.339200550844386
5 0.316516811570287
7 0.315056347957495
};
\end{axis}

\end{tikzpicture}
    \vspace*{-1.5em}
    \caption{$\textit{GED}_{s}$ on DBPedia}
  \end{subfigure}
  \begin{subfigure}[c]{0.24\linewidth}
\begin{tikzpicture}[
scale=0.5,
every axis/.style={
                    legend pos=south west, 
                    legend style={
                font=\small},
                    legend columns=2, 
                }
]

\definecolor{lightblue161201244}{RGB}{161,201,244}
\definecolor{lightgreen141229161}{RGB}{141,229,161}
\definecolor{lightgrey203}{RGB}{203,203,203}
\definecolor{lightsalmon255159155}{RGB}{255,159,155}
\definecolor{lightsalmon255180130}{RGB}{255,180,130}
\definecolor{thistle208187255}{RGB}{208,187,255}
\definecolor{whitesmoke240}{RGB}{240,240,240}

\begin{axis}[
axis line style={whitesmoke240},
height=1.5\textwidth,
tick align=outside,
tick pos=left,
width=2\textwidth,
x grid style={lightgrey203},
xlabel={\(\displaystyle k\)},
xmajorgrids,
xmin=1.75, xmax=7.25,
xtick style={color=black},
xtick={1,2,3,4,5,6,7,8},
xticklabels={
  \(\displaystyle {1}\),
  \(\displaystyle {2}\),
  \(\displaystyle {3}\),
  \(\displaystyle {4}\),
  \(\displaystyle {5}\),
  \(\displaystyle {6}\),
  \(\displaystyle {7}\),
  \(\displaystyle {8}\)
},
y grid style={lightgrey203},
ylabel={Mean \(\displaystyle GED_{s}\)},
ymajorgrids,
ymin=0, ymax=1,
ytick style={color=black},
ytick={0,0.2,0.4,0.6,0.8,1},
yticklabels={
  \(\displaystyle {0.0}\),
  \(\displaystyle {0.2}\),
  \(\displaystyle {0.4}\),
  \(\displaystyle {0.6}\),
  \(\displaystyle {0.8}\),
  \(\displaystyle {1.0}\)
}
]
\addplot [thick, lightblue161201244, mark=x, mark size=5, mark options={solid,fill opacity=0}]
table {%
2 0.618342546455094
3 0.50966946887514
5 0.455133553297926
7 0.436699062630227
};
\addplot [thick, lightblue161201244, dashed, mark=x, mark size=5, mark options={solid,fill opacity=0}]
table {%
2 0.341075710859884
3 0.298762531328321
5 0.293939242477254
7 0.412116830537883
};
\addplot [thick, lightsalmon255180130, mark=x, mark size=5, mark options={solid,fill opacity=0}]
table {%
2 0.503357314148681
3 0.531537777240424
5 0.471981968017774
7 0.429322344322344
};
\addplot [thick, lightsalmon255180130, dashed, mark=x, mark size=5, mark options={solid,fill opacity=0}]
table {%
2 0.29652278177458
3 0.313776963776964
5 0.292673506948973
7 0.400963948332369
};
\addplot [thick, lightgreen141229161, mark=x, mark size=5, mark options={solid,fill opacity=0}]
table {%
2 0.72757793764988
3 0.674595687331536
5 0.634039298072614
7 0.77647000861636
};
\addplot [thick, lightgreen141229161, dashed, mark=x, mark size=5, mark options={solid,fill opacity=0}]
table {%
2 0.385131894484412
3 0.527228327228327
5 0.500340756967263
7 0.668607068607069
};
\addplot [thick, lightsalmon255159155, mark=x, mark size=5, mark options={solid,fill opacity=0}]
table {%
2 0.678486997635934
3 0.597867063492064
5 0.586308203991131
7 0.693349668349668
};
\addplot [thick, lightsalmon255159155, dashed, mark=x, mark size=5, mark options={solid,fill opacity=0}]
table {%
2 0.343802769334684
3 0.410515873015873
5 0.384860803153486
7 0.541196416196416
};
\addplot [thick, thistle208187255, mark=x, mark size=5, mark options={solid,fill opacity=0}]
table {%
2 0.604255319148936
3 0.534285714285714
5 0.609613997113997
7 0.658299193496562
};
\addplot [thick, thistle208187255, dashed, mark=x, mark size=5, mark options={solid,fill opacity=0}]
table {%
2 0.334751773049645
3 0.489464285714286
5 0.447306397306397
7 0.58342972816657
};
\end{axis}

\end{tikzpicture}
    \vspace*{-1.5em}
    \caption{$\textit{GED}_{s}$ on Yago}
  \end{subfigure}
  \begin{subfigure}[c]{0.24\linewidth}
\begin{tikzpicture}[
scale=0.5,
every axis/.style={
                    legend pos=south west, 
                    legend style={
                font=\small},
                    legend columns=2, 
                }
]

\definecolor{lightblue161201244}{RGB}{161,201,244}
\definecolor{lightgreen141229161}{RGB}{141,229,161}
\definecolor{lightgrey203}{RGB}{203,203,203}
\definecolor{lightsalmon255159155}{RGB}{255,159,155}
\definecolor{lightsalmon255180130}{RGB}{255,180,130}
\definecolor{thistle208187255}{RGB}{208,187,255}
\definecolor{whitesmoke240}{RGB}{240,240,240}

\begin{axis}[
axis line style={whitesmoke240},
height=1.5\textwidth,
tick align=outside,
tick pos=left,
width=2\textwidth,
x grid style={lightgrey203},
xlabel={\(\displaystyle k\)},
xmajorgrids,
xmin=1.75, xmax=7.25,
xtick style={color=black},
xtick={1,2,3,4,5,6,7,8},
xticklabels={
  \(\displaystyle {1}\),
  \(\displaystyle {2}\),
  \(\displaystyle {3}\),
  \(\displaystyle {4}\),
  \(\displaystyle {5}\),
  \(\displaystyle {6}\),
  \(\displaystyle {7}\),
  \(\displaystyle {8}\)
},
y grid style={lightgrey203},
ylabel={Mean \(\displaystyle GED_{s}\)},
ymajorgrids,
ymin=0, ymax=1,
ytick style={color=black},
ytick={0,0.2,0.4,0.6,0.8,1},
yticklabels={
  \(\displaystyle {0.0}\),
  \(\displaystyle {0.2}\),
  \(\displaystyle {0.4}\),
  \(\displaystyle {0.6}\),
  \(\displaystyle {0.8}\),
  \(\displaystyle {1.0}\)
}
]
\addplot [thick, lightblue161201244, mark=x, mark size=5, mark options={solid,fill opacity=0}]
table {%
2 0.431985972683647
3 0.585987696514012
5 0.458801344184585
7 0.576923076923077
};
\addplot [thick, lightblue161201244, dashed, mark=x, mark size=5, mark options={solid,fill opacity=0}]
table {%
2 0.271650055370986
3 0.204868231184021
5 0.228658943936722
7 0.384615384615385
};
\addplot [thick, lightsalmon255180130, mark=x, mark size=5, mark options={solid,fill opacity=0}]
table {%
2 0.348702600526979
3 0.589603159189487
5 0.54787296037296
7 0.31486076647367
};
\addplot [thick, lightsalmon255180130, dashed, mark=x, mark size=5, mark options={solid,fill opacity=0}]
table {%
2 0.0583979328165375
3 0.0741854636591479
5 0.181313131313131
7 0.2
};
\addplot [thick, lightgreen141229161, mark=x, mark size=5, mark options={solid,fill opacity=0}]
table {%
2 0.48671977971876
3 0.605111176702086
5 0.552105248774043
7 0.400234448382426
};
\addplot [thick, lightgreen141229161, dashed, mark=x, mark size=5, mark options={solid,fill opacity=0}]
table {%
2 0.233333333333333
3 0.218903318903319
5 0.26984126984127
7 0.0242424242424242
};
\addplot [thick, lightsalmon255159155, mark=x, mark size=5, mark options={solid,fill opacity=0}]
table {%
2 0.45905325443787
3 0.626257734732311
5 0.490532544378698
7 0.769230769230769
};
\addplot [thick, lightsalmon255159155, dashed, mark=x, mark size=5, mark options={solid,fill opacity=0}]
table {%
2 0.0707692307692308
3 0.088512241054614
5 0.112820512820513
7 0.461538461538462
};
\addplot [thick, thistle208187255, mark=x, mark size=5, mark options={solid,fill opacity=0}]
table {%
2 0.385227272727273
3 0.460640301318267
5 0.426332288401254
7 0.576923076923077
};
\addplot [thick, thistle208187255, dashed, mark=x, mark size=5, mark options={solid,fill opacity=0}]
table {%
2 0.103535353535354
3 0.139924670433145
5 0.120689655172414
7 0.230769230769231
};
\end{axis}

\end{tikzpicture}
    \vspace*{-1.5em}
    \caption{$\textit{GED}_{s}$ on \acrshort{bto}}
  \end{subfigure}
  \begin{subfigure}[c]{0.24\linewidth}
\begin{tikzpicture}[
scale=0.5,
every axis/.style={
                    legend pos=south west, 
                    legend style={
                font=\small},
                    legend columns=2, 
                }
]

\definecolor{lightblue161201244}{RGB}{161,201,244}
\definecolor{lightgreen141229161}{RGB}{141,229,161}
\definecolor{lightgrey203}{RGB}{203,203,203}
\definecolor{lightsalmon255159155}{RGB}{255,159,155}
\definecolor{lightsalmon255180130}{RGB}{255,180,130}
\definecolor{thistle208187255}{RGB}{208,187,255}
\definecolor{whitesmoke240}{RGB}{240,240,240}

\begin{axis}[
axis line style={whitesmoke240},
height=1.5\textwidth,
tick align=outside,
tick pos=left,
width=2\textwidth,
x grid style={lightgrey203},
xlabel={\(\displaystyle k\)},
xmajorgrids,
xmin=1.75, xmax=7.25,
xtick style={color=black},
xtick={1,2,3,4,5,6,7,8},
xticklabels={
  \(\displaystyle {1}\),
  \(\displaystyle {2}\),
  \(\displaystyle {3}\),
  \(\displaystyle {4}\),
  \(\displaystyle {5}\),
  \(\displaystyle {6}\),
  \(\displaystyle {7}\),
  \(\displaystyle {8}\)
},
y grid style={lightgrey203},
ylabel={Mean \(\displaystyle GED_{s}\)},
ymajorgrids,
ymin=0, ymax=1,
ytick style={color=black},
ytick={0,0.2,0.4,0.6,0.8,1},
yticklabels={
  \(\displaystyle {0.0}\),
  \(\displaystyle {0.2}\),
  \(\displaystyle {0.4}\),
  \(\displaystyle {0.6}\),
  \(\displaystyle {0.8}\),
  \(\displaystyle {1.0}\)
}
]
\addplot [thick, lightblue161201244, mark=x, mark size=5, mark options={solid,fill opacity=0}]
table {%
2 0.613123340873321
3 0.595207935264807
5 0.523956546598056
7 0.379490701606086
};
\addplot [thick, lightblue161201244, dashed, mark=x, mark size=5, mark options={solid,fill opacity=0}]
table {%
2 0.199776471770143
3 0.188228905597327
5 0.133259821939067
7 0.253085376162299
};
\addplot [thick, lightsalmon255180130, mark=x, mark size=5, mark options={solid,fill opacity=0}]
table {%
2 0.562992654185302
3 0.494962286030581
5 0.539392049124375
7 0.49010989010989
};
\addplot [thick, lightsalmon255180130, dashed, mark=x, mark size=5, mark options={solid,fill opacity=0}]
table {%
2 0.124902665468703
3 0.071752858849633
5 0.173295454545455
7 0.274725274725275
};
\addplot [thick, lightgreen141229161, mark=x, mark size=5, mark options={solid,fill opacity=0}]
table {%
2 0.584252786220871
3 0.574950094324036
5 0.52594696969697
7 0.534216034591974
};
\addplot [thick, lightgreen141229161, dashed, mark=x, mark size=5, mark options={solid,fill opacity=0}]
table {%
2 0.16422619047619
3 0.214345965262196
5 0.281089561023772
7 0.363500784929356
};
\addplot [thick, lightsalmon255159155, mark=x, mark size=5, mark options={solid,fill opacity=0}]
table {%
2 0.591024249004705
3 0.527737180514958
5 0.489696414696415
7 0.413008558527426
};
\addplot [thick, lightsalmon255159155, dashed, mark=x, mark size=5, mark options={solid,fill opacity=0}]
table {%
2 0.140825190010858
3 0.138929588929589
5 0.160521736098659
7 0.0888997532865457
};
\addplot [thick, thistle208187255, mark=x, mark size=5, mark options={solid,fill opacity=0}]
table {%
2 0.543516022761306
3 0.517466550319953
5 0.50137741046832
7 0.585724230520611
};
\addplot [thick, thistle208187255, dashed, mark=x, mark size=5, mark options={solid,fill opacity=0}]
table {%
2 0.177253668763103
3 0.130536856976648
5 0.113228001944579
7 0.209861932938856
};
\end{axis}

\end{tikzpicture}
    \vspace*{-1.5em}
    \caption{$\textit{GED}_{s}$ on UniProt}
  \end{subfigure}

  \caption{$F_{1,\textit{node}}$, $F_{1,\textit{rel}}$, and $\textit{GED}_{s}$ on four different ontologies, comparing different models and node amounts $k$, and the \textit{raw} and \textit{aligned} (constrained) output.}
  \label{fig:compare_onto_model_scores}
\end{figure*}
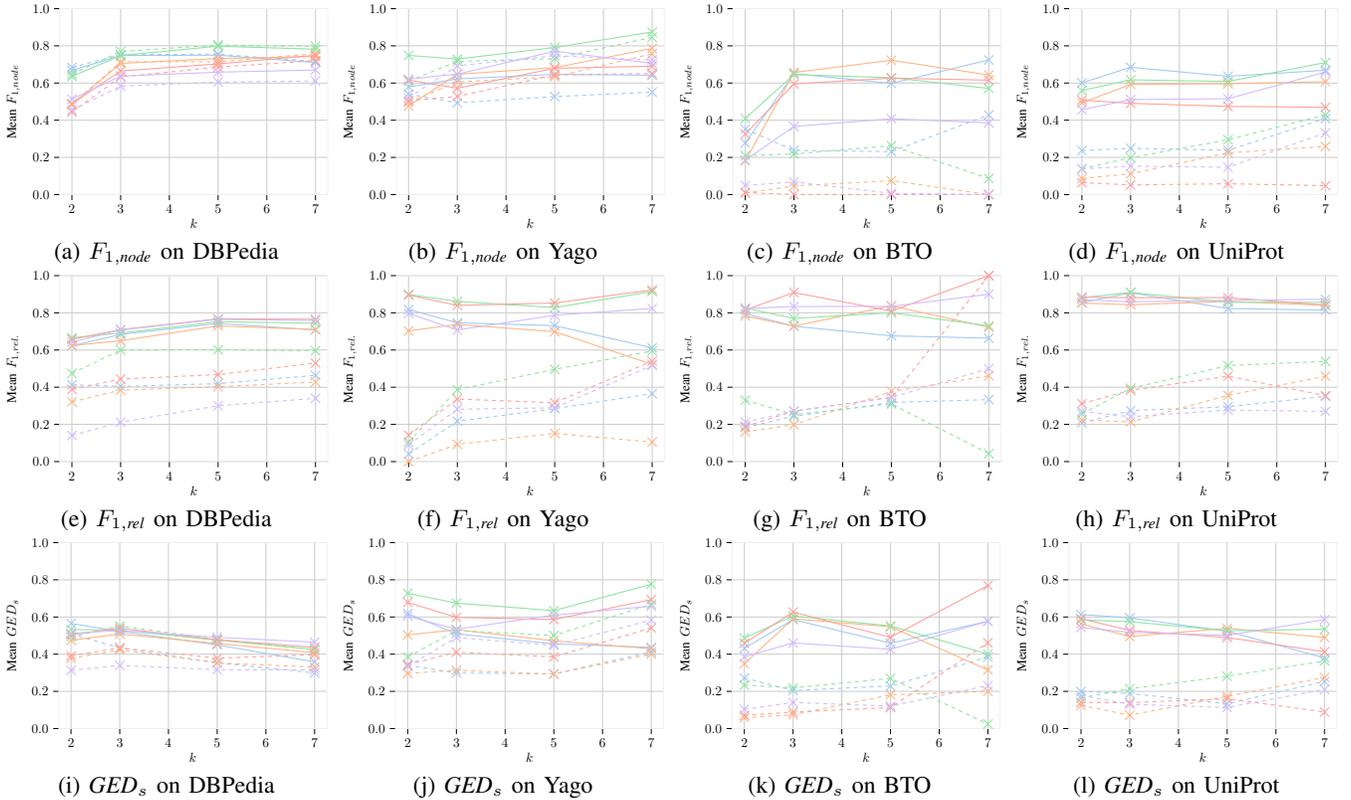

Our evaluation in \cref{fig:compare_onto_model_scores} shows that we can faithfully recover the prototype graph from the \gls{nl} query. While we cannot achieve a perfect recovery of all nodes and relations in all settings, we accomplish a $F_1$ score on the node retrieval on DBpedia across all models of approximately $0.7$ throughout the different levels of complexity. The correct relations are retrieved at an even higher $F_1$ score of roughly $0.8$ for the various ontologies and settings. 
Similarly, the \gls{ged} score $\textit{GED}_{s}$ shows that the graph structure can be recovered quite well for most of the smaller graph sizes and drops slightly for the larger, more complex queries, demonstrating that our approach is most useful for smaller graphs. 
Our evaluation, furthermore, shows that using a larger model is only slightly beneficial for our use case. The system performs similarly across all model sizes and types, suggesting that even smaller models can reconstruct the prototype graph quite well due to the constraints we place on the output.

The evaluation also shows that the constraint and alignment step to the ontology, including our corrections, is essential to retrieve the correct graph. This is evident by the ablation study performed by comparing the \textit{raw} output of the \gls{lm} to the \textit{aligned} outputs of our fill pipelines. The system achieves indeed higher scores in the \textit{aligned} results, and can be observed for almost all combinations of query complexities, \gls{lm} models, and ontologies.

\begin{table}
  \centering
  \caption{Comparison of query generation methods for Hermes 3B on DBpedia.}
  \label{tab:compare_gen_query}
  \begin{tabular}{lllrr}
    \toprule
                          &                               &         & Mean         & Mean      \\
    $k$                   & Query Origin                  & Stage   & $F_{1,node}$ & $GED_{s}$ \\
    \midrule
    \multirow[t]{6}{*}{3} & \multirow[t]{2}{*}{human}     & aligned & 0.63         & 0.40      \\
                          &                               & raw     & 0.52         & 0.24      \\
    \cline{2-5}
                          & \multirow[t]{2}{*}{llama}     & aligned & 0.71         & 0.46      \\
                          &                               & raw     & 0.68         & 0.36      \\
    \cline{2-5}
                          & \multirow[t]{2}{*}{templated} & aligned & 0.81         & 0.57      \\
                          &                               & raw     & 0.79         & 0.54      \\
    \cline{1-5} \cline{2-5}
    \multirow[t]{6}{*}{5} & \multirow[t]{2}{*}{human}     & aligned & 0.69         & 0.22      \\
                          &                               & raw     & 0.90         & 0.27      \\
    \cline{2-5}
                          & \multirow[t]{2}{*}{llama}     & aligned & 0.66         & 0.34      \\
                          &                               & raw     & 0.67         & 0.25      \\
    \cline{2-5}
                          & \multirow[t]{2}{*}{templated} & aligned & 0.83         & 0.48      \\
                          &                               & raw     & 0.74         & 0.45      \\
    \bottomrule
  \end{tabular}
\end{table}

Finally, we compare our different query generation methods in \cref{tab:compare_gen_query} by evaluating two query complexities $k=\{3,5\}$ and comparing the scores. The \gls{lm}-based generation method performs similarly to the human-generated queries. This comparative evaluation, additionally, shows that the template queries perform better than the compared \gls{lm}-generated ones. These results indicate that the \glspl{lm} perform better with simpler formatted queries, such as the template-generated queries.

\subsection{User Study}

We also inspect the preliminary results of our user study shown in \cref{tab:task_results}, which involved $n=11$ participants. We observe that our average correct completion rate (\enquote{Success Rate}) is high throughout the increasingly complex tasks, compared to the study on \textit{RDFExplorer}~\cite{Vargas2019RDF}. They report that their correctness decreases, possibly due to time constraints. Within the survey of our tool, however, no user took longer than the 20-minute time frame, and we observed lower success rates only for the first and third tasks. These tasks might have been more challenging because the \gls{lm} did not directly provide the correct prototype graph. The most complex task, however, had the highest correctness rate compared to the others in our study. These results, in contrast to the observations of the \textit{RDFExplorer}, indicate that our system performs consistently across varying levels of difficulty and primarily relies on the \gls{lm} output.
The evaluation of our \gls{sus} questionnaire resulted in a score of 71.4.

\begin{table}
  \centering
  \caption{Task results for the OnSET user study on DBpedia with $n=11$.}
  \label{tab:task_results}
  \begin{tabular}{lrl}
    \toprule
    Task   & Success Rate & Time (mm:ss) \\
    \midrule
    Task 0 & 0.73         & 03:46        \\
    Task 1 & 0.91         & 01:12        \\
    Task 2 & 0.73         & 04:49        \\
    Task 3 & 1.00         & 02:05        \\
    \bottomrule
  \end{tabular}
\end{table}

  \section{Conclusion}
We introduce a novel querying system in this paper to aid users in exploring ontologies and generating queries from \gls{nl}. The interface uses only \gls{nl} as the input, providing the users with a fuzzy interface to explore an otherwise strict system of classes and links. We achieve this translation from fuzzy input to the strict notion using a query processing pipeline that heavily utilizes \glspl{lm} for generating the prototype graph, performing semantic similarity search, and constrained graph generation. We provide a prototype graph as an intermediate output from this pipeline. This prototype graph can be refined through a node-based editor to better reflect the users' information needs if the \gls{lm} failed to map the \gls{nl} query completely, or the user has a more specific information need.

We evaluate our system on a set of synthetic queries with varying levels of complexity. These should reflect users' varying information needs and queries while evaluating our system for more complex tasks. This evaluation demonstrates that the extracted prototype graphs accurately represent the sampled graphs, particularly for smaller graphs. This performance level, especially at the smaller graph sizes, should provide a robust system that adheres to the users' initial interests. We demonstrate that this level of performance can be achieved with smaller models. In contrast, existing systems struggle with these smaller models and require significantly more powerful, and thus more expensive, models. They, furthermore, often require iterative systems to retrieve syntactically correct queries, whereas our system can provide valid queries using our two-step retrieval process. We furthermore validate our approach by comparing the synthetic queries generated by a \gls{lm} and template system from the sampled graphs to a small set of manually written queries. These evaluations show that the \gls{lm}-generated queries do indeed adhere to the manually written queries, thereby strengthening our query evaluation approach.

We additionally conducted a small-scale user study to assess our systems' performance with respect to usability and task completion time, where we observed high rates of correctness and fast completion times compared to existing systems. This suggests that combining a precise and efficient \gls{lm} system with a visual query editor can reduce the time spent on creating queries while improving the correctness of the results compared to prior works.

\section{Future Work}
The introduced system builds a prototype graph based on \gls{nl} input, enabling users to explore and query knowledge graphs effectively. Future versions could, however, enhance the robustness of retrieving the prototype graph from \gls{nl} queries for larger and more complex queries by utilizing more sophisticated models or fine-tuning smaller, more specialized models for the graph retrieval task at hand, thereby achieving even more precise initial graph responses.

Another avenue for further improvement in this system is the addition of constraints to the graph generated by the \glspl{lm}. While we can constrain the properties (e.g., the age, name, or height of a person) within the user interface, we do not provide a way for the \gls{lm} to add this to the generated query. These, however, are an integral part of the \gls{kg} and the information they contain. We intend to extend the \gls{lm} output and graph to include these filters on the edges and formalize these structures to get a constrained output, including these constraints.
The user study will also be extended to include more users, with additional questionnaires and possibly more challenging tasks, to facilitate a more comprehensive comparison with other systems using visual query builders.

  \bibliographystyle{IEEEtran}
  \bibliography{IEEEabrv,references}

  \appendix

\subsection{Graph sampling}

\begin{balgorithm}[t]
  \caption{Prototype graph $G_{p,s}$ sampling from the ontology using instance counts within the \gls{kg}.}\label{alg:sampling}
  \label{alg:graph_sampling}
  \begin{algorithmic}
    \Require classes $\mathcal{C}$, links $\mathcal{L}$, objects $\mathcal{O}$, predicates $\mathcal{P}$, samples $k > 0$, depth $d>0$, maximum of nodes $m_{\text{nodes}}>0$

    \Function{downgrade node}{$n_s$}
    \State $\mathcal{C}_{sub} \gets \text{subtypes of $n_s$ up to depth $d$}$
    \State $Pr_{node}(n_s)=\frac{|\{n_{I,h,i}\in  \mathcal{0}|n_s=\mathrm{typeof}(n_{I,h,i})\}|}{\sum_{n_s'\in \mathcal{C}_{sub}}  |\{n_{I,h,i}\in  \mathcal{0}|n_s'=\mathrm{typeof}(n_{I,h,i})\}|}$\\
    \Return $n\sim Pr_{node}(n_s)$
    \EndFunction

    \Function{downgrade link}{$l=(n_{p,i},n_{p,j},l_{ij})$}
    \State $n_{p,i,\text{sub}}\gets \Call{downgrade node}{n_{p,i}}$
    \State $n_{p,j,\text{sub}}\gets  \Call{downgrade node}{n_{p,j}}$\\
    \Return $l=(n_{p,i,\text{sub}},n_{p,j,\text{sub}},l_{ij})$
    \EndFunction

    \State $\mathcal{L}_{\mathrm{cand.}} \gets \text{top $k$ links}~\mathcal{L}$ \Comment{By instance count}
    \State $P_{link}(l\in\mathcal{L}_{\mathrm{cand.}})=\frac{|\{e_{I,h,ij}\in  \mathcal{P}|l=\mathrm{typeof}(e_{I,h,ij})\}|}{\sum_{l'\in\mathcal{L}_{\mathrm{cand.}}}|\{e_{I,h,ij}\in  \mathcal{P}|l'=\mathrm{typeof}(e_{I,h,ij})\}|}$
    \State $l_{\mathrm{sel}}\sim P(l\in\mathcal{L}_{\mathrm{cand.}})$

    \State $l_{sub}\gets\Call{downgrade link}{l}$
    \State $E_{p,s}\gets\{l_sub\}$
    \State $N_{p,s}\gets\{n_{p,0,\text{sub}},n_{p,1,\text{sub}}\}$

    \While{$m_{\text{nodes}} >|N_{p,s}|$}
    \State $side\sim \mathcal{U}\{left,right\}$
    \State $n_{sel}\sim  \mathcal{U}(N_p)$
    \State $\mathcal{L}_{\mathrm{cand., next}}\gets \text{top $k$ links attaching to $side$}$
    \State $l_{new}\sim P_{link}(l\in \mathcal{L}_{\mathrm{cand., next}})$
    \State $l_{new,sel}\gets\Call{downgrade link}{l_{new}}$
    \State $E_{p,s}\gets E_{p,s} \cup \{l_{new,sel}\}$
    \State $N_{p,s}\gets N_{p,s} \cup  \{ \text{added node of $l_{new,sel}$} \}$
    \EndWhile
    \State \textbf{output} $G_{p,s}=(N_{p,s}, E_{p,s})$
  \end{algorithmic}
\end{balgorithm}
\label{ssec:sampling}
We sample the ontology for queries using the procedure in \cref{alg:graph_sampling}. The outlined algorithm should provide a broad variety of graph structures, classes, and relations to create a diverse set of queries for our evaluation, while being faithful to the interest of the users based on the probabilistic sampling.

\subsection{Query Generation}
\label{ssec:querygen}

We generate the SPARQL queries from our prototype graphs $G_p$ directly by iterating over all links and nodes. Each link results in the link type and left and right nodes; each node adds the node as a \verb|a| class.

\cref{lst:example_query} shows a resulting query for an example query -- the same query as in \cref{fig:teaser,fig:user_interface}.

\subsection{Reproducibility}
\label{ssec:repro}

To ensure the reproducibility of our evaluation and enable further experiments on our user interface, we provide our code and parameters on \href{https://github.com/Dakantz/OnSET}{\texttt{github.com/Dakantz/OnSET}}. We performed the experiments on the smaller models ($\leq$32B parameters) on a system with an RTX 4090, and the large-scale \gls{lm} experiments on our university cluster on multiple Quadro RTX 8000s.

\begin{lstlisting}[language=SPARQL, label=lst:example_query, caption={Example SPARQL query generated from the input \enquote{a person and the child of a person have the alma mater of the same university}.}]
SELECT DISTINCT ?person_1 ?person_2 ?university_1 WHERE {
    ?person_1 <http://dbpedia.org/ontology/child> ?person_2.
    ?person_1 <http://dbpedia.org/ontology/almaMater> ?university_1.
    ?person_2 <http://dbpedia.org/ontology/almaMater> ?university_1.
    ?person_1 a <http://dbpedia.org/ontology/Person>.
    ?person_2 a <http://dbpedia.org/ontology/Person>.
    ?university_1 a <http://dbpedia.org/ontology/University>.
}
\end{lstlisting}
\end{document}